\documentclass[aps,prb,superscriptaddress,amsfonts,amsmath,amssymb,floatfix,reprint,longbibliography]{revtex4-2}

\usepackage{booktabs}
\usepackage[dvipsnames]{xcolor}
\usepackage[colorlinks=true, urlcolor=blue, linkcolor=blue, citecolor=blue, pdftex]{hyperref}
\usepackage{multirow}
\usepackage{natbib}
\usepackage{siunitx}
\DeclareSIUnit\erg{erg}

\usepackage{tikz}
\usepackage[version=4]{mhchem}
\usepackage{textgreek}

\usepackage{braket}
\newcommand{\chemSP}{Cu$_6$Al(SO$_4$)(OH)$_{12}$Cl$\cdot$3H$_2$O}

\definecolor{hamCol1}{HTML}{1D00E9}
\definecolor{hamCol2}{HTML}{F60014}
\definecolor{hamCol3}{HTML}{F29200}
\definecolor{hamCol4}{HTML}{0087EF}
\definecolor{hamCol5}{HTML}{93C1D0}
\usepackage{pdfpages}
\makeatletter
\AtBeginDocument{\let\LS@rot\@undefined}
\makeatother

\begin{document}


\title{Tensor network analysis of the maple-leaf antiferromagnet spangolite}

\author{Philipp Schmoll}
\email{philipp.schmoll@fu-berlin.de}
\affiliation{Freie Universität Berlin, Dahlem Center for Complex Quantum Systems and Institut für Theoretische Physik,  Arnimallee 14, 14195 Berlin, Germany}

\author{Harald O. Jeschke}
\email{jeschke@okayama-u.ac.jp}
\affiliation{Research Institute for Interdisciplinary Science, Okayama University, Okayama 700-8530, Japan}
\affiliation{Department of Physics and Quantum Centre of Excellence for Diamond and
Emergent Materials (QuCenDiEM), Indian Institute of Technology Madras, Chennai 600036, India}

\author{Yasir Iqbal}
\email{yiqbal@physics.iitm.ac.in}
\affiliation{Department of Physics and Quantum Centre of Excellence for Diamond and
Emergent Materials (QuCenDiEM), Indian Institute of Technology Madras, Chennai 600036, India}


\begin{abstract}
Spangolite (\chemSP) is a hydroxy-hydrated copper sulfate mineral with a one-seventh depleted triangular lattice of Cu$^{2+}$ ions in each layer. Experimental measurements revealed a non-magnetic ground state at $T \sim \SI{8}{\kelvin}$ with magnetic properties dominated by dimerization. We propose a spatially anisotropic Heisenberg model for the Cu$^{2+}$ spin-$1/2$ degrees of freedom on this geometrically frustrated and effectively two-dimensional maple-leaf lattice, featuring five symmetry inequivalent couplings with ferromagnetic bonds on hexagons and antiferromagnetic triangular bonds. The validity of the proposed Hamiltonian is demonstrated by state-of-the-art tensor network calculations, which can assess both the nature of the ground state as well as low-temperature thermodynamics, including the effects of a magnetic field. We provide theoretical support for a picture of a non-trivially correlated dimer ground state, which accounts for the appreciable reduction of the magnetic moment at high temperatures observed in experiment, thereby resolving a long-standing puzzle. We predict the static spin structure factor as well as the emergence of magnetisation plateaus at high values of an external magnetic field, explore the nature of the quantum states in them, and study their melting with increasing temperature.
\end{abstract}

\maketitle

\section*{Introduction}

\begin{figure*}[htb]
    \includegraphics[width=\textwidth]{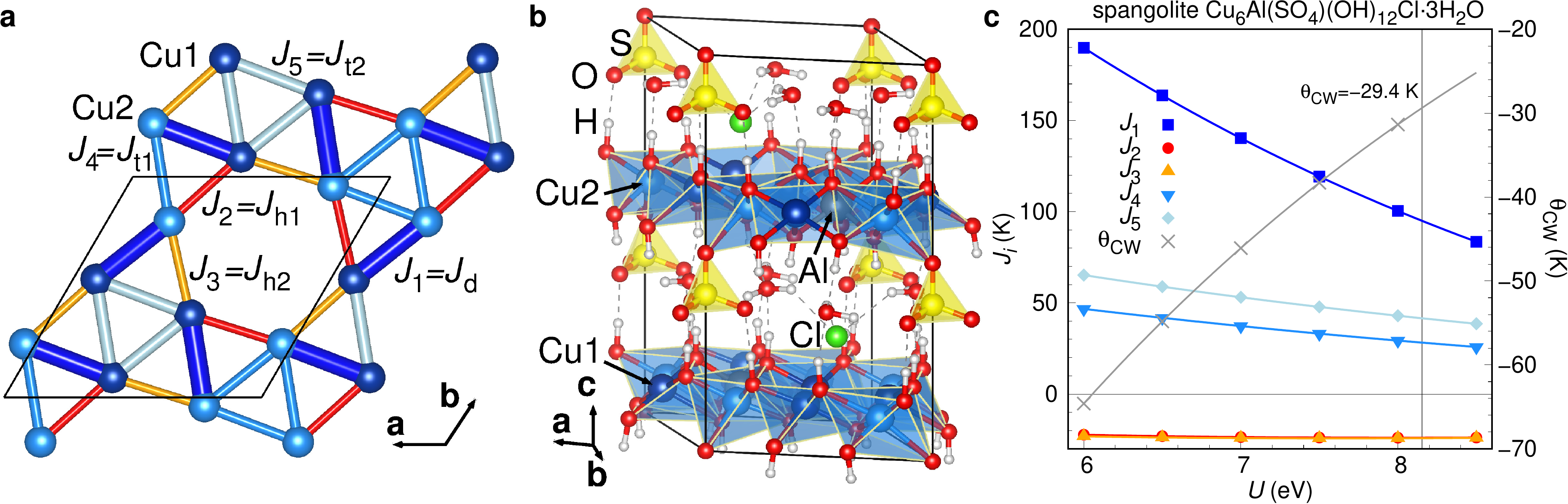}
    \caption{{\bf Structure and Heisenberg Hamiltonian parameters of spangolite.} {\bf a} Distorted maple-leaf lattice realized in spangolite. The bonds represent the exchange interactions $J_i$, with thickness scaled in proportion to their values. $J_1$ to $J_5$ are named with ascending Cu-Cu distance, while $J_{\rm t1}$, $J_{\rm t2}$ (triangles), $J_{\rm d}$ (dimer) and $J_{\rm h1}$, $J_{\rm h2}$ (hexagon) are used in the literature. {\bf b} Crystal structure of spangolite {\chemSP}~\cite{Fennell2011} with DFT relaxed hydrogen positions. {\bf c} Five exchange interactions constituting the maple-leaf lattice, determined by DFT energy mapping, as function of on-site interaction strength $U$. The vertical line indicates the $U$ value for which the couplings match the experimental Curie-Weiss temperature, and for this $U$ we have antiferromagnetic $J_1=95.0(2)$\,K, $J_4=28.2(1)$\,K and $J_5=41.6(2)$\,K and ferromagnetic $J_2=-23.9(1)$\,K and $J_3=-24.2(1)$\,K. }
    \label{fig:Fig1}
\end{figure*}

The kagome lattice antiferromagnet is the embodiment of high geometric frustration in two dimensions~\cite{Huse-1992}. However, a number of other lattices based on triangular motifs come close if we consider suppression of ordered moment~\cite{Farnell2011,Farnell-2014}. Examples of other highly frustrated Archimedean lattices are the star, bounce, trellis and maple-leaf lattices. A common feature of these lattices is that material realizations are extremely rare~\cite{Zheng-2014}; metal-organic approximate versions exist for the star~\cite{Zheng2007} and trellis lattice~\cite{Yamaguchi2018}. Concerning the maple-leaf lattice (MLL) (see Fig.~\ref{fig:Fig1}\,{\bf a})~\cite{Betts-1995}, the situation is slightly more promising~\cite{Norman2018,Inosov2018} as a number of minerals with quantum spins like bluebellite \ce{Cu6IO3(OH)10Cl}~\cite{Mills2014}, mojaveite \ce{Cu6TeO4(OH)Cl}~\cite{Mills2014}, fuettererite \ce{Pb3Cu6TeO6(OH)7Cl5}~\cite{Kampf2013}, sabelliite (Cu,Zn)$_2$Zn[(As,Sb)O$_4$](OH)$_3$~\cite{Olmi1995} and spangolite {\chemSP}~\cite{Penfield1890,Miers-1893,Frondel-1949,Hawthorne1993} have been found; besides, some semi-classical maple-leaf lattice antiferromagnets like MgMn$_3$O$_7$$\cdot$3H$_2$O~\cite{Haraguchi2018} and \ce{Na2Mn3O7}~\cite{Venkatesh2020,Saha-2023} are known. Here, we focus on spangolite, which has been characterised magnetically~\cite{Fennell2011} with evidence of a non-magnetic ground state. However, the nature of the singlet ground state remains a riddle with various speculative scenarios of either isolated or interacting dimers and trimers having been discussed in Ref.~\cite{Farnell2011}, but none found to be in complete agreement with observed magnetic features. In this manuscript, we address this long-standing issue by first employing ab-initio density functional theory (DFT) calculations to reliably ascertain the magnetic interactions, which are shown in Fig.~\ref{fig:Fig1}\,{\bf a}, and found to be antiferromagnetic on the dimer ($J_{1}$) and triangular ($J_{4}$ and $J_{5}$) bonds, and ferromagnetic on the hexagons ($J_{2}$ and $J_{3}$). The resulting magnetic Hamiltonian is analysed employing state-of-the-art tensor network (TN) simulations based on {infinite projected entangled simplex states} (iPESS)~\cite{Xie2014} and {infinite projected entangled simplex operators} (iPESO)~\cite{Schmoll2022} ans\"atze, to assess both its ground state as well as finite magnetic field and finite-temperature behaviour. 

Our results lend support to a picture of a dimerized ground state characterised by strong singlet formation on the dimer bonds ($J_{1}$ in Fig.~\ref{fig:Fig1}\,{\bf a}), which can be ascribed to the presence of a large antiferromagnetic coupling on these bonds. Since these singlets are coupled via appreciable ferromagnetic correlations on the hexagons, they cannot be viewed as being isolated. The ground state is thus composed of correlated dimers and can be viewed as a dressed version of the exact dimer product state~\cite{Ghosh-2022}. The behaviour of the magnetic susceptibility with temperature is found to be in qualitative agreement with experiment. For a quantitative match of the calculated susceptibility, a smaller spin gap would be required which would consequently lead to a higher maximum; it is plausible that once an experimental low temperature crystal structure becomes available, redetermination of the Hamiltonian could yield a slightly smaller $J_1$ and larger frustration, yielding a smaller spin gap. The finding of a substantially reduced effective magnetic moment at high temperatures (in agreement with experiment) compared to that expected of six $S=1/2$ spins lends support to a ground state composed of non-trivially correlated dimers. This is indeed confirmed by the obtained ground state spin correlation profile, which features significant spin-spin correlations on the triangular and hexagonal bonds. We do not see any evidence of the formation of strongly bound clusters or trimers, which would require invoking additional orbital moments as speculated in Ref.~\cite{Fennell2011}. Under the application of a magnetic field, the resulting magnetisation curve displays plateaus at $1/3$ and $2/3$ of the saturation magnetisation characterised by a translationally invariant pattern of spin-spin correlations. The temperature evolution of the magnetisation curve shows a relatively faster melting of the $1/3$ compared to the $2/3$ plateau.

\begin{table*}
\begin{tabular}{c|c||c|c|c||c|c|c||c}
role & name & \multicolumn{3}{c||}{spangolite} & \multicolumn{3}{c||}{bluebellite}  & name\\
 & (this work) & $J_i$\,(K) & $J_i/J_{\rm max}$ & $d_{\rm Cu-Cu}$\,(\AA)  & $J_i$\,(K) & $J_i/J_{\rm max}$& $d_{\rm Cu-Cu}$\,(\AA) & Ref.~\cite{Makuta2021} \\\hline
dimer       & $J_1$ & {\bf 95.0(2)} & 1     & 3.005 & -120.8 & -0.82 & 2.992  & $J_{\rm d}$ \\
hexagon 1   & $J_2$ & {\bf -23.9(1)} & -0.252 & 3.107 & 88.6   & 0.60  & 3.000 & $J_{\rm h1}$ \\
hexagon 2   & $J_3$ & {\bf -24.2(1)} & -0.254 & 3.110 & -93.7  & -0.63 & 3.165 & $J_{\rm h2}$ \\
triangle 1  & $J_4$ & {\bf 28.2(1)}  & 0.297  & 3.213 & 147.6  & 1     & 3.287 & $J_{\rm t1}$ \\
triangle 2  & $J_5$ & {\bf 41.6(2)}  & 0.437  & 3.216 & 61.3   & 0.42  & 3.453 & $J_{\rm t2}$ \\\hline
\end{tabular}
    \caption{{\bf Exchange interactions of spangolite obtained by DFT energy mapping and compared to bluebellite}. The spangolite couplings (given in bold face) are interpolated to $U=\SI{8.15}{\electronvolt}$ so that they match the Curie-Weiss temperature \mbox{$\theta_{\rm CW}=\SI{-29.4}{\kelvin}$} (see Methods). The Cu-Cu distances are given to identify the exchange path. The Hamiltonian of bluebellite \ce{Cu6IO3(OH)10Cl} from Ref.~\cite{Ghosh-2023_bluebellite} is given for comparison.}    \label{tab:couplings}
\end{table*}

\begin{figure}[htb]
    \includegraphics[width = \columnwidth]{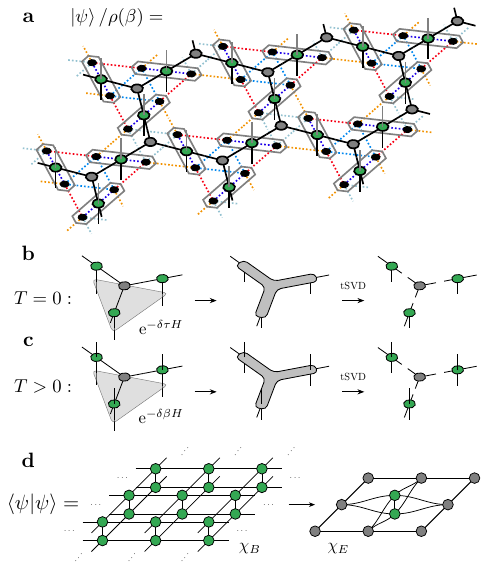}
    \caption{{\bf Tensor network setup for the simulations of spangolite.} {\bf a} Infinite projected entangled simplex state and operator {\it ansatz} for the simulations of spangolite. A coarse-graining of the two spins on $J_1$ bonds results in a regular kagome lattice, on which the tensor network is defined. Here, green tensors reside on the coarse-grained sites, encompassing the physical degrees of freedom, while gray tensors mediate quantum correlations through purely virtual bonds, linking the former. While quantum states $\ket{\psi}$ have one physical index (represented by the black vertical lines), thermal states $\rho(\beta)$ need two physical indices (both black and gray vertical lines). {\bf b} \& {\bf c} Simple update step for the iPESS ground state ($T = 0$) and iPESO thermal state ($T > 0$). The imaginary time evolution gate is first absorbed into a triangle configuration to evolve the states. Applying a higher-order singular value decomposition (SVD) with truncation (tSVD) to the bulk bond dimension  $\chi_{\rm B}$ recovers the individual tensors. {\bf d} Coarse-graining the six spins in the elementary unit cell results in a square lattice TN, whose contraction is approximated by fixed-point environment tensors shown in gray, using a corner transfer matrix renormalization group procedure. The unavoidable approximations are controlled by an environment bond dimension $\chi_{\rm E}$.}
    \label{fig:Fig2}
\end{figure}

\begin{figure*}
    \includegraphics[width = 0.98\textwidth]{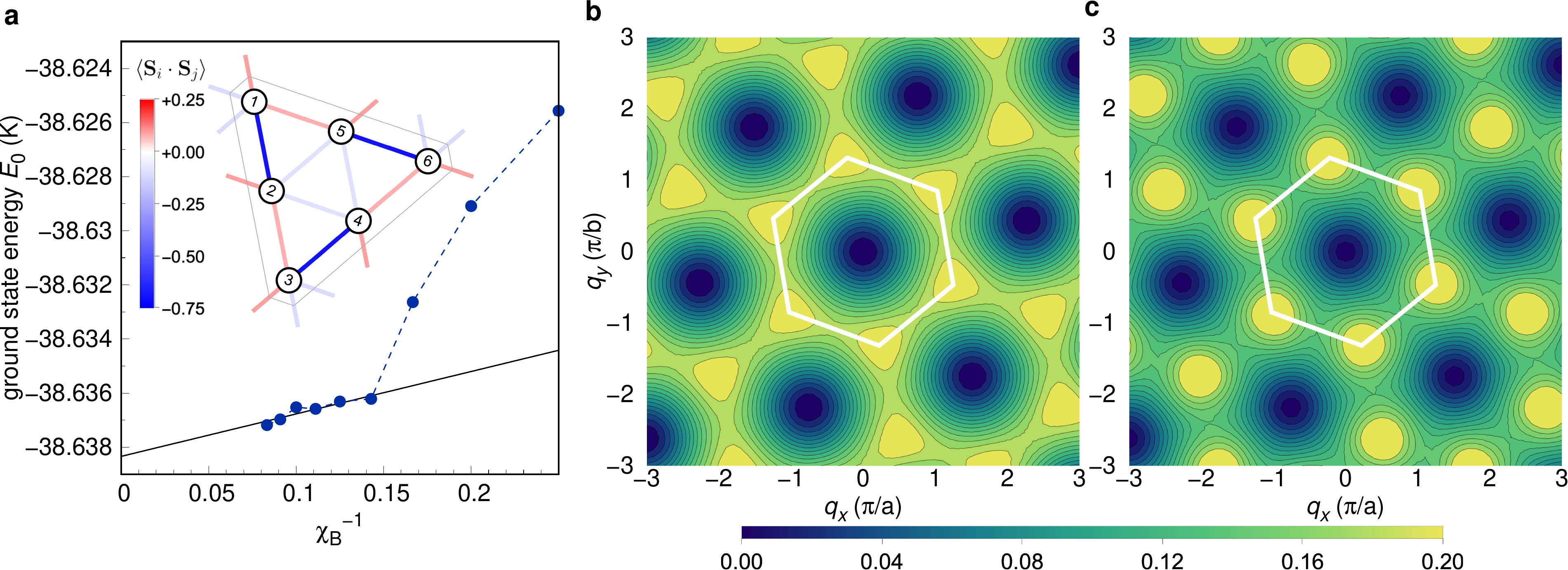}
    \caption{{\bf Ground state of spangolite.} {\bf a} Ground state energy vs. inverse bond dimension $\chi_{\rm B}$ for the ground state of the spangolite Hamiltonian without magnetic field. The inset shows the spatial pattern of nearest-neighbour spin-spin correlations, revealing the strong dimerization. {\bf b} The equal-time spin structure factor $S(\bf q)$ for an exact dimer state on $J_1$ bonds~\cite{Ghosh-2023_bluebellite,Gresista-2023} and {\bf c} for the ground state of the spangolite Hamiltonian at bond dimension $\chi_{\rm B} = 8$. The extended Brillouin zone is marked by a white hexagon.}
    \label{fig:Fig3}
\end{figure*}

\section*{Results}

{\bf Heisenberg Hamiltonian}

We begin our study of spangolite by determining the parameters of the Heisenberg Hamiltonian
\begin{equation}
    H=\sum_{i<j} J_{ij} {\bf S}_i\cdot {\bf S}_j
    \label{eq:H}
\end{equation}
using the density functional theory based energy mapping technique. This approach has proven instrumental for the understanding of various Cu$^{2+}$ based minerals like centennialite~\cite{Iida2020}, henmilite~\cite{Yamamoto2021}, birchite~\cite{Fujihara2022}, kapellasite~\cite{Iqbal-2015} as well as PbCuTe$_{2}$O$_{6}$ (idealized coloalite~\cite{Powell1994})~\cite{Chillal-2020}. The first step is to establish the best possible crystal structure, including all hydrogen positions. For this purpose, we use the structure determined by Fennell {\it et al.}~\cite{Fennell2011}, adding to it the missing H3 position from Hawthorne {\it et al.}~\cite{Hawthorne1993}. We then carefully relax all six hydrogen positions using the full potential local orbital (FPLO) code~\cite{Koepernik1999} and the generalized gradient approximation (GGA)~\cite{Perdew1996} exchange correlation functional. The resulting crystal structure is shown in Fig.~\ref{fig:Fig1}\,{\bf b}. The structure is characterised by Cu$^{2+}$ maple-leaf layers where Al$^{3+}$ ions fill the centers of the hexagons. The metal hydroxide layers are well separated by sulfate groups and water molecules, indicating a high level of two-dimensionality of the material.

We now proceed to extract the Heisenberg Hamiltonian parameters of Eq.~\eqref{eq:H} by DFT energy mapping (see Method section for details). We can resolve the five nearest-neighbour exchange interactions which make up the distorted maple-leaf lattice of spangolite. We also determine seven longer range couplings which turn out to be less than one percent of the largest coupling and which we ignore in our subsequent analysis. The values of the first five exchange interactions are plotted in Fig.~\ref{fig:Fig1}\,{\bf c} as function of the on-site interaction strength $U$. The relevant value of $U$, determined by demanding that the mean-field estimate of the Curie-Weiss temperature matches the experimental value of $\theta_{\rm CW}=-29.4$\,K~\cite{Fennell2011} is marked by a vertical line (note that we redetermined $\theta_{\rm CW}$ from the experimental data as explained in the Method section). The resulting Hamiltonian parameters, which are the basis for our further investigation are given in Table~\ref{tab:couplings}. Spangolite is found to be characterised by a unique network of three antiferromagnetic and two ferromagnetic couplings shown in Fig.~\ref{fig:Fig1}\,{\bf a}. The largest coupling, antiferromagnetic $J_1$, defines dimers. The two weaker antiferromagnetic couplings $J_4=0.297 J_1$ and $J_5=0.437 J_1$ define triangles, and the two ferromagnetic couplings $J_2=-0.252 J_1$ and $J_3=-0.254 J_1$ form the hexagons of the maple-leaf lattice. 

In order to understand the unique features of spangolite, it is instructive to compare it to bluebellite which has been investigated in Refs~\cite{Haraguchi2021,Makuta2021,Ghosh-2023_bluebellite} (see Table~\ref{tab:couplings}). The dimer coupling $J_1\equiv J_{\rm d}$ which is strong for both spangolite and bluebellite is nearly equal in magnitude but {opposite} in sign, antiferromagnetic and ferromagnetic, respectively. This implies that a dimer singlet picture is the natural starting point to understanding the physics of spangolite as all other couplings are clearly subdominant. In contrast, in bluebellite, strong ferromagnetic $J_{\rm d}$ bonds compete with even stronger antiferromagnetic triangles, which leads to a very different starting point based on antiferromagnetically frustrated triangles.

The two sets of hexagon bonds $J_{2}$, $J_{3}$ ($J_{\rm h1}$, $J_{\rm h2}$) are both ferromagnetic and of nearly equal magnitude in spangolite, while in bluebellite they are nearly equal in magnitude but of opposite signs. This leads to inherently strongly frustrated hexagons in bluebellite but unfrustrated hexagonal units in spangolite. Hence, this points to completely different frustration mechanisms at play in the two materials. Furthermore, the hexagonal couplings in bluebellite are $\sim 60\%$ of the strongest coupling and cannot simply be treated as a perturbation of a parent order determined by $J_{\rm t1}$ and $J_{\rm d}$ couplings. On the other hand, for spangolite, the hexagonal couplings are only $\sim 20\%$ of the dimer coupling and do not alter the picture of weakly correlated dimers.
In the following, we will analyse the properties of the spangolite Hamiltonian using state-of-the-art tensor network techniques.\\

{\bf Tensor network ansatz}

The study of the spangolite Hamiltonian is based on numerical tensor network simulations in the thermodynamic limit. TNs are efficient representations of quantum many-body systems, that encode the probability amplitudes of a (thermal) quantum state as a contraction of a network of local tensors. The tensors are interconnected by auxiliary, virtual indices as shown in Fig.~\ref{fig:Fig2}\,{\bf a}, whose maximal dimension is called the \textit{bond dimension} $\chi_{\rm B}$ of the TN (bulk bond dimension). It is a control parameter that can be systematically increased to improve the accuracy of the {\it ansatz}. Tuning the bond dimension changes the number of variational parameters in the TN and thereby the amount of quantum entanglement that can be captured. Tensor networks offer efficient numerical simulations with only a polynomial scaling in the number of particles, thus overcoming the exponential barrier by targeting the low-entanglement sector of the full Hilbert space~\cite{Eisert2010,Orus2014}. In our study, we employ {infinite projected entangled simplex states} (iPESS)~\cite{Xie2014} and {infinite projected entangled simplex operators} (iPESO)~\cite{Schmoll2022} for the simulation of ground states and thermal states, respectively. In this context, the
TN is used as an {\it ansatz} for the full many-body system, consisting of a unit cell of different tensors that generates a translationally invariant quantum state. The simple update steps for the iPESS and iPESO networks visualized in Figs.~\ref{fig:Fig2}\,{\bf b} and {\bf c} are described Supplementary Notes 2 and 3, together with the coarse-graining scheme shown in Fig.~\ref{fig:Fig2}\,{\bf d}, and other details of the TN implementation.\\

{\bf Ground state without magnetic field}

The ground state for the spangolite Hamiltonian can be represented with a single geometrical unit cell of six spins. Due to the strong anti\-ferro\-magnetic interaction on $J_1$ bonds, it is dominated by a dimerization of the connected spins. The remaining antiferromagnetic interaction terms contribute only weakly to the structure, however, there exist nonnegligible ferromagnetic correlations on the hexagonal bonds. Due to the chosen coarse-graining, the two spins on $J_1$ bonds are treated exactly as a single tensor site, and entanglement with neighbouring sites is only weak. The chosen TN {\it ansatz} is therefore ideal and the simple update algorithm leads to accurate results. More sophisticated procedures, such as variational optimisation~\cite{Naumann2023} are limited to smaller bond dimensions and not superior here. The ground state energy vs. the inverse iPESS bond dimension is shown in Fig.~\ref{fig:Fig3}\,{\bf a}. Quantum correlations are already seen to be well captured with an {\it ansatz} of $\chi_{\rm B} = 8$, and a larger bond dimension does not substantially decrease the energy further. A fit of the six largest data points reveals a good estimate for the infinite bond dimension limit $(\chi_{\rm B} \rightarrow \infty)$, for which we extrapolate to an energy of
\begin{align}
    E_0/J_1 = \num{-0.4066(4)}.
\end{align}
The energy is found to be lower compared to that of an exact dimer product state (on $J_{1}$ bonds), which has an energy per site of $E/J_{1}=-0.375$. The spatial pattern of nearest-neighbour spin-spin correlations $\langle \mathbf S_i \cdot \mathbf S_j \rangle$ reveals the expected strong dimerization of the ground state on $J_1$ bonds. It is shown as the inset of Fig.~\ref{fig:Fig3}\,{\bf a}. The spins on the hexagons show weak ferro\-magnetic alignment, while the spins on triangle configurations are nearly uncorrelated. Thus, the ground state cannot be characterised by isolated singlet formations on the dimer bonds alone, but instead these dimers are correlated (due to ferromagnetically correlated hexagonal bonds) forming a correlated dimer liquid with a lower energy. 

In order to further characterize the ground state, we compute the equal-time spin structure factor
\begin{align}
    \begin{aligned}
        S({\bf q}) &= \sum_{i,j} \sum_{m,n} e^{\dot\iota {\bf q}\cdot({\bf R}_i - {\bf R}_j)}  e^{\dot\iota {\bf q}\cdot({\bf b}_m - {\bf b}_n)} \\
        & \times \left\langle {\bf S}({\bf R}_i + {\bf b}_m) \cdot {\bf S}({\bf R}_j + {\bf b}_n) \right\rangle.
    \end{aligned}
    \label{eq:staticSpinStructureFactor}
\end{align}
Here, $(i,j)$ denotes the summation over unit cells and $(m,n)$ the additional summation over the six-site basis of the maple-leaf lattice. By exploiting translational invariance, one sum over unit cells can be removed, and the structure factor can be computed by a CTMRG resummation scheme~\cite{Corboz2016, Ponsioen2020,Ponsioen2023}, accounting for the appropriate phase factors and spin operators while absorbing tensors into the environment tensors. This scheme typically needs large environment bond dimensions $\chi_{\rm E}$ to reach convergence. However, due to the large gap in our system (cf. the width of the zero-magnetisation plateau below), the calculations of the structure factor can be converged with moderate $\chi_{\rm E}$. In order to have a defined reference, we show the structure factor for the spangolite ground state in comparison to an exact dimer state on $J_1$ bonds in Fig.~\ref{fig:Fig3}\,{\bf b} and {\bf c}.
While the exact dimer state is representable with an iPESS at bond dimension $\chi_{\rm B} = 1$, i.e., a product state of coarse-grained dimers, the structure factor for spangolite is computed at $\chi_{\rm B} = 8$ to balance accuracy and the computational cost of the CTMRG routine. The similarity in the structure factor for the spangolite ground state with the exact dimer one confirms the strong dimerization, however, it is slightly smeared compared to the exact dimer state.\\

{\bf Thermal state and heat capacity}

\begin{figure}[ht]
    \includegraphics[width = 0.95\columnwidth]{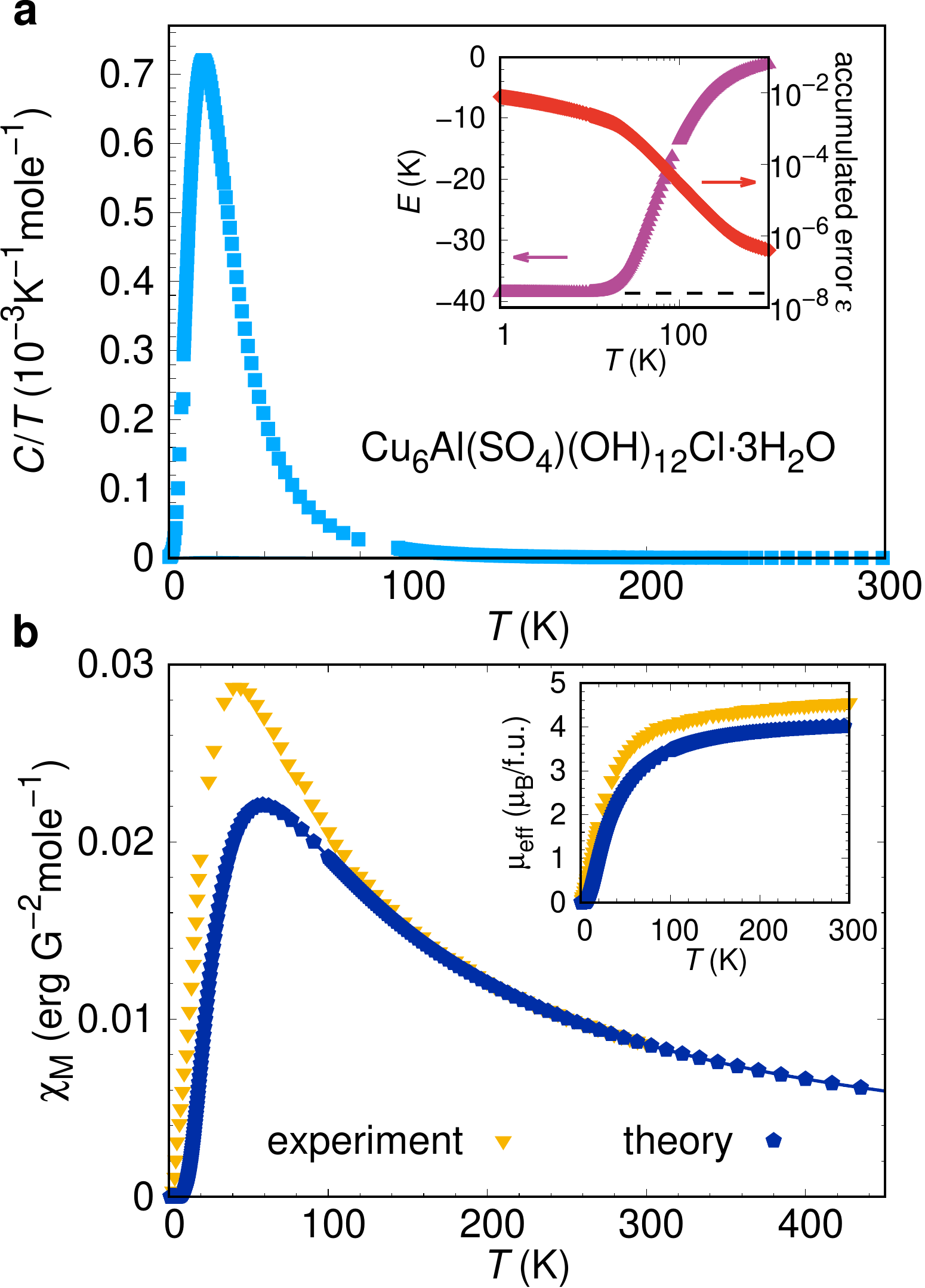}
    \caption{{\bf Thermodynamic properties of spangolite.} {\bf a} Specific heat divided by temperature showing a pronounced peak at $T \sim \SI{20}{\kelvin}$. The inset shows the thermal state energy, that converges to the ground state energy for low temperatures, alongside the accumulated truncation error of the simple update cooling procedure, that stays below $\varepsilon \sim 10^{-2}$ throughout the full range of temperatures. {\bf b} Magnetic susceptibility over temperature. The high-temperature regime ($T > \SI{1000}{\kelvin}$ for the simulation, $T > \SI{250}{\kelvin}$ for the experiment) is fitted with a Curie–Weiss law to extract the CW temperature in Table~\ref{tab:fitParametersCW}. The inset shows the effective magnetic moment. The theoretical susceptibility data have been scaled by a factor of $\sim 0.68$ to match the high-temperature tail as explained in the text.}
    \label{fig:Fig4}
\end{figure}

Before investigating magnetic properties of the spangolite Hamiltonian, we compute thermal state properties without a magnetic field using iPESO simulations at a large bulk bond dimension $\chi_{\rm B} = 24$ with mean-field environments~\cite{Schmoll2022}. Thermal states are obtained by successively cooling an infinite-temperature state with infinitesimal steps $\delta\beta = 10^{-4}$ down to low temperatures. From the thermal state energy we can compute the heat capacity $C = \partial U / \partial T$. Results are shown in Fig.~\ref{fig:Fig4}\,{\bf a}. The heat capacity $C/T$ features a pronounced peak at $T \sim \SI{20}{\kelvin}$.

In the inset we show the thermal state energy alongside the truncation error in the simple update cooling procedure. The thermal state energy converges to the ground state energy of Fig.~\ref{fig:Fig3}\,{\bf a} for low temperatures, indicating that the procedure does not get stuck in local minima during the cooling. The truncation error can be used to probe the accuracy of the simulations. It stays below $\varepsilon \sim 10^{-2}$ down to $T = \SI{1}{\kelvin}$, which is a good indication that the chosen bond dimension for the thermal state simulations is sufficiently high.\\

{\bf Magnetic susceptibility}

To further characterize the spangolite Hamiltonian, we computed the variation of the magnetic susceptibility with temperature using the iPESO thermal state algorithm. To this end, we choose again a bulk bond dimension of $\chi_{\rm B} = 24$ and an accurate infinitesimal temperature step of $\delta\beta = 10^{-4}$, leading to a maximally accessible temperature of $1/\delta\beta = 10^{4}$\,K. Expectation values are computed using the mean-field environment, which is reasonable due to the high bond dimension and temperature, and low entanglement in the system. Fortunately, we can directly compare our theoretical simulations of the model Hamiltonian to experimental measurements on spangolite. The magnetic susceptibility is computed from the magnetisation along the direction of the field according to
\begin{align}
    \chi_{\rm M}(T) = \frac{\partial m_z(T)}{\partial h_z} \bigg\vert_{h_z \rightarrow 0}.
\end{align}
Our results, alongside the data extracted from Ref.~\cite{Fennell2011} are shown in Fig.~\ref{fig:Fig4}\,{\bf b}. The main feature of a sharp peak and the high-temperature tail are correctly recovered by our tensor network simulations. A numerical fit of the high-temperature regime (\mbox{$T > \SI{1000}{\kelvin}$} for the simulation, \mbox{$T > \SI{250}{\kelvin}$} for the experiment) with a Curie-Weiss law
\begin{align}
    \chi_{\rm M}(T) = \frac{C}{T - \theta_\text{CW}}
\end{align}
is used to extract the Curie-Weiss temperature $\theta_\text{CW}$. The fit results in parameters given in Table~\ref{tab:fitParametersCW}.
\begin{table}[ht]
    \centering
    \begin{tabular}{c | c c}
         & experiment & simulation \\
        \hline
        $C$\,$\big(\frac{\text{erg}\,K}{G^2 \text{mole}}\big)$ & $2.85$ & $4.23$ \\
        $\theta_\text{CW}$\,(K) & $-29.4$ & $-31.5$ \\
        $\mu_\text{eff}$\,$(\mu_{\rm B}/{\rm f.u.})$ & $4.76$ & $5.82$ \\
    \end{tabular}
    \caption{Parameters of the numerical fit of the Curie-Weiss law of the magnetic susceptibility. Experimental data from Ref.~\protect\cite{Fennell2011} have been fitted as explained in the Method section. The effective magnetic moment $\mu_\text{eff}$ is computed according to Eq.~\eqref{eq:effectiveMagneticMoment} in the units of $\mu_{\rm B}$ per formula unit.}
    \label{tab:fitParametersCW}
\end{table}
The simulations show a peak at a slightly higher temperature ($T=59$\,K) compared to experiment ($T=44$\,K) and a quicker decline of the magnetic susceptibility, indicating a non-magnetic ground state already at higher temperatures. From the Curie-Weiss constant $C$ we can compute the effective magnetic moment
\begin{align}
    \mu_\text{eff} = \sqrt{\frac{3 k_B C}{ N_A \mu_{\rm B}^2}}.
    \label{eq:effectiveMagneticMoment}
\end{align}
The extracted value from the TN simulations is slightly larger compared to the experimentally estimated one, see Table~\ref{tab:fitParametersCW}. However, our calculations capture the substantial reduction in $\mu_\text{eff}$ compared to the expected value for the spangolite formula unit, which has \mbox{$\mu_\text{eff} = \num{10.39}\,\mu_{\rm B}/{\rm f.u.}$} for the six spin-$1/2$s~\cite{Fennell2011}. This reflects the fact that the ground state cannot be viewed as either being composed of isolated dimers or explained by a model of dimers interacting with a mean-field~\cite{Haraldsen-2005,Singh-2007,Castro-2010} and confirms the description of the ground state obtained from TN simulations. In order to match the high-temperature tail of the susceptibility to the experimental data, we need to scale the Weiss constant $C$ by a factor of $\sim 0.68$. The experimental finding is that the expected effective moment of $\mu_{\rm eff}=10.39\,\mu_{\rm B}$ for a unit cell of six Cu $S=1/2$ is reduced to $\mu_{\rm eff}=4.75\,\mu_{\rm B}$ which is only 46{\%} of the expected value and corresponds to a reduced spin value of $S=0.14$. Theoretically, we find a reduction of the effective moment to $\mu_{\rm eff}=5.82\,\mu_{\rm B}$ which is 56{\%} of the expected value and corresponds to a reduced spin value of $S=0.20$. Thus, our simulations capture the major part of the moment reduction but fall a bit short. The missing  reduction in the magnetic moment could be related to enhanced frustration upon introduction of inter-layer couplings which are inevitably present in the real material. 
The fact that our calculated susceptibility vanishes at a somewhat higher temperature than in experiment indicates an overestimation of the formation of dimers, which could result from a slightly too high value for $J_1$ in the model Hamiltonian. The iPESO {\it ansatz} itself is biased towards the formation of dimers on $J_1$ bonds as well, however, the large bulk bond dimension $\chi_{\rm B} = 24$ should allow sufficient correlations to neighbouring sites to counteract this bias. Using $C = \chi_{\rm M} T$ in Eq.~\eqref{eq:effectiveMagneticMoment}, we can plot the effective magnetic moment as function of temperature as shown in the inset of Fig.~\ref{fig:Fig4}\,{\bf b}. While neither experiment nor simulation data for the effective moment completely saturate at room temperature,  our result approaches a slightly lower value compared to the experiment, in agreement with the preceding discussion.\\

\begin{figure*}[htb]
    \centering
    \includegraphics[width = 0.95\textwidth]{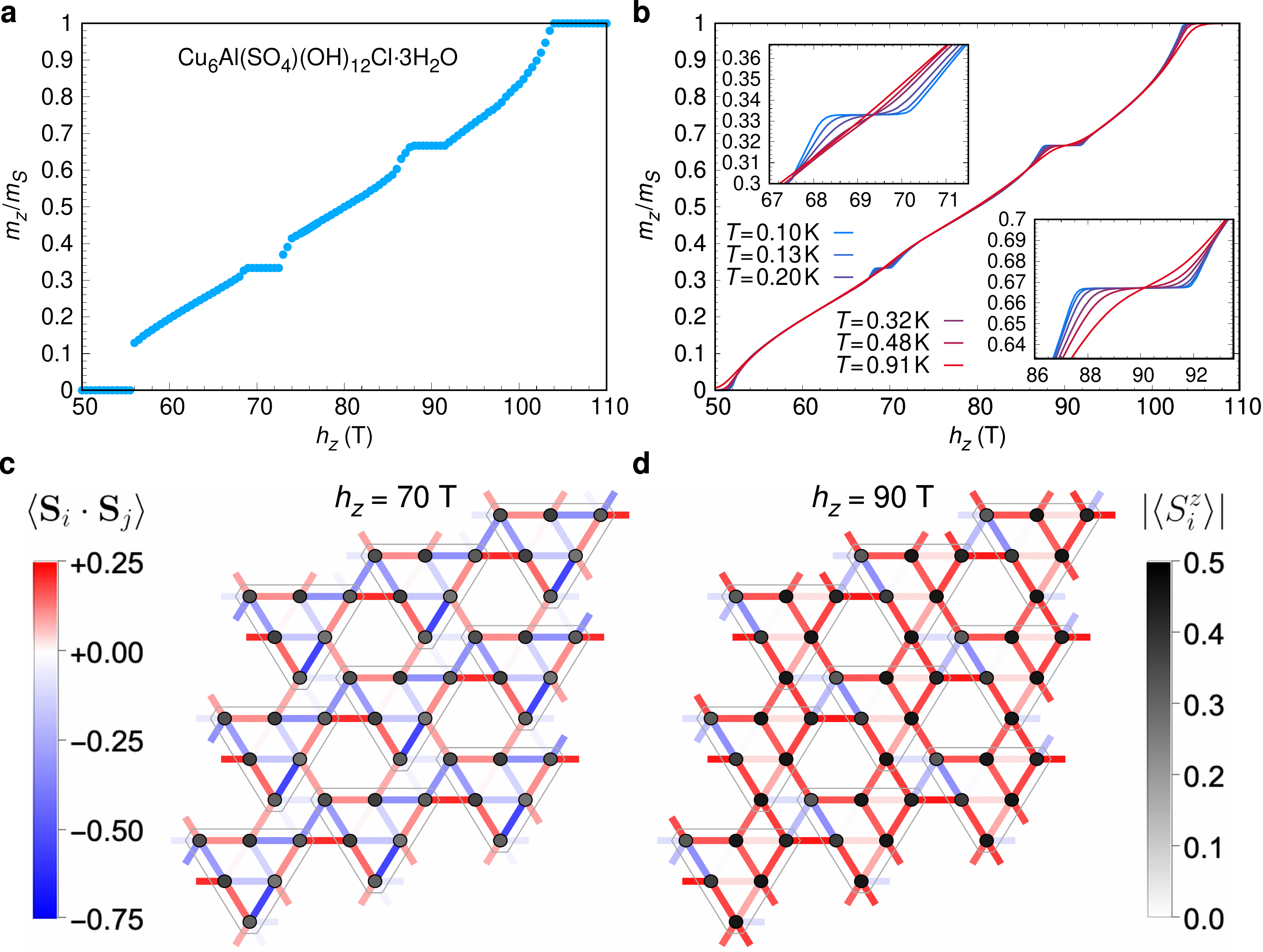}
    \caption{{\bf Magnetisation curves of spangolite.} {\bf a} Zero temperature magnetisation of the spangolite Hamiltonian over the magnetic field $h_z$ at $\chi_{\rm B} = 8$, using CTMRG environments with $\chi_{\rm E} = 64$. Below $h_z \sim \SI{56}{\tesla}$ the strong antiferromagnetic $J_1$ interaction leads to a vanishing magnetisation. Two plateaus at $1/3$ and $2/3$ of the saturation $m_{\rm S} = 1/2$ are found. {\bf b} Magnetisation curve of spangolite at finite temperatures. The insets show the melting of the $m_z/m_{\rm S} = 1/3$ and $2/3$ magnetisation plateaus as an effect of increasing temperature. {\bf c}-{\bf d} Nearest-neighbour spin-spin correlations and local $S^z$ expectation values for the $1/3$ and $2/3$ magnetisation plateaus, corresponding to values of {\bf c} $h_z = 70$\,T and {\bf d} $h_z = 90$\,T, respectively.}
    \label{fig:Fig5}
\end{figure*}

{\bf Magnetisation plateaus}

Finally, we study the spangolite Hamiltonian in an external magnetic field. Due to the strong antiferromagnetic interaction on $J_1$ bonds, the system retains a non-magnetic ground state up until $h_z \sim \SI{56}{\tesla}$. In the $T = \SI{0}{\kelvin}$ simulations we have used unit cells of six, twelve and eighteen spins in order to capture more complicated ground state structures. 
Upon tuning the magnetic field, we find the emergence of magnetisation plateaus at values of $m_z/m_{\rm S} = 1/3$ and $2/3$ of the saturation value $m_{\rm S} = 1/2$. The full magnetisation curve is shown in Fig.~\ref{fig:Fig5}\,{\bf a}. It is interesting to note that we do not find the $1/6$, $2/9$ and $2/7$ plateaus observed in the magnetisation curve of the exact dimer product ground state~\cite{Ghosh-2023_mf}. This provides further evidence that the ground state of the spangolite Hamiltonian is not formed by isolated singlets but more appropriately viewed as a correlated dimer state due to the presence of ferromagnetic correlations on the hexagonal bonds connecting the dimers. 

The spin-spin correlations in both the $1/3$ and $2/3$ plateaus are translationally invariant and thus representable within a six site unit cell. Those plateaus satisfy the conventional condition for a magnetisation plateau to emerge, namely, $N_{\rm u}S(1-m_z/m_{\rm s})\in\mathbb{Z}$~\cite{Affleck-1988,Oshikawa-1997}, where $N_{\rm u}$ is the number of sites in the unit cell for the ground state and $S$ is the spin quantum number which, in the present case, is $S=1/2$. For the aforementioned quantities, we obtain values of $2$ and $1$ for the $1/3$ and $2/3$ plateaus, respectively. The spatial pattern of spin-spin correlations and $z$-component of the magnetisation is shown in Fig.~\ref{fig:Fig5}\,{\bf c}-{\bf d}.

Finally, we analyse the stability of the observed plateaus with respect to temperature. To this end, we simulate the spangolite Hamiltonian in a field over a temperature range $T \, \in \, [\SI{0.1}{\kelvin}, \SI{1000}{\kelvin}]$ at a fixed bond dimension of $\chi_{\rm B} = 24$ using mean-field environments to compute expectation values. Here we use a slightly larger temperature step $\delta\beta = 10^{-3}$ to achieve more efficient cooling. The unit cell is set to six spins. While the infinite-temperature states are fully disordered without any magnetisation, the simple update cooling procedure recovers the two plateaus at low temperatures. A set of fixed temperature slices is shown in Fig.~\ref{fig:Fig5}\,{\bf b}.
In the inset we show the melting of both plateaus. Noticeably, they melt rather symmetrically from the low and high field limits, respectively. Due to the smaller width, the $1/3$ magnetisation plateau is more susceptible to thermal effects and starts melting sooner than the $2/3$ plateau.

\section*{Discussion}
Employing a combination of density functional theory based energy mapping and tensor network simulations, we investigated the ground and thermal state properties in the presence of a magnetic field of a distorted maple-leaf lattice as realized in spangolite (\chemSP). The DFT Hamiltonian features the largest coupling, antiferromagnetic $J_{1}$ on the dimer bonds, while the subleading antiferromagnetic couplings, $J_{4}=0.297 J_{1}$ and $J_{5}=0.437 J_{1}$ define the triangles. A noteworthy aspect of the Hamiltonian is the presence of substantial ferromagnetic interactions, $J_{2}=-0.252 J_{1}$ and $J_{3}=-0.254 J_{1}$ on the hexagons. The pattern of spin-spin correlations in the ground state consequently features strong singlet formation on the dimer bonds, and sizeable ferromagnetic correlations on the hexagonal bonds which connect the dimers. Interestingly, the spins on the triangles are nearly uncorrelated. Thus, our picture of a non-trivially correlated dimer ground state accounts for the appreciable reduction in magnetic moment at high temperatures observed in experiment, thereby resolving a long-standing puzzle. It also effectively rules out a scenario of a ground state composed of either isolated dimers or dimers interacting with a mean-field, and dispenses with the need of invoking more complicated mechanisms which have been speculated about earlier. We find a good agreement with the experimentally observed behaviour of the magnetic susceptibility. The magnetisation curve of spangolite does not show the presence of $1/6$, $2/9$, and $2/7$ plateaus which are expected for an exact dimer product configuration. This lends further evidence of a correlated dimer ground state. We find that the $1/3$ and $2/3$ plateau states are fully translation invariant. With increasing temperature, the $2/3$ plateau appears more robust compared to the $1/3$ plateau.

Much of the recent theoretical efforts in search of exotic non-magnetic states on the maple-leaf lattice have focused on the purely antiferromagnetic model~\cite{Gresista-2023,Beck-2024,Gembe-2024}. However, lately the explorations of models with mixed ferro-antiferro couplings (with antiferromagnetically coupled hexagons and ferromagnetic coupling on triangles and dimer bonds) have unveiled a variety of dimer orders encompassing an island of putative quantum spin liquid behaviour~\cite{ghosh2024spin}. In a similar vein, there could be a putative quantum spin liquid lurking around the corner proximate to the observed correlated dimer ground state. To this end, it would be worthwhile to perform a thorough investigation of the mixed ferro-antiferro parameter space of the spangolite Hamiltonian, which features ferromagnetically coupled hexagons and antiferromagnetic coupling on triangles and dimer bonds. This might also enable us to ascertain whether the observed dimerized phase could potentially be viewed as an instability of proximate quantum spin liquids, which have lately been classified in Ref.~\cite{sonnenschein2024}.

Our study of spangolite is the second detailed elucidation of the Hamiltonian of a maple-leaf material, the first being the recent investigation of bluebellite~\cite{Ghosh-2023_bluebellite}. Both minerals share the distortion pattern of the ideal maple-leaf lattice with five inequivalent nearest-neighbour bonds and the fact that the Hamiltonian is dominantly antiferromagnetic, but modified by two ferromagnetic exchange interactions. As the resultant physics, spin singlet formation versus valence bond solid ground state for spangolite and bluebellite, respectively, vary considerably, it is a worthwhile endeavor to understand the other known copper minerals with maple-leaf lattice, mojaveite, fuetterite and sabelliite, as well. Their maple-leaf lattices are also distorted, promising new patterns of antiferromagnetic and ferromagnetic bonds. The example of bluebellite~\cite{Haraguchi2018} suggests that synthesis of the minerals is possible and can lead to a reduction of the lattice distortions. A next step could be efforts to synthesize material variations, for example by replacing chlorine by bromine and thus to further extend the richness of quantum spin systems with maple-leaf structure.

\section*{Methods}
\subsection*{Density functional theory based energy mapping}
We use all electron density functional theory calculations with a full potential local orbital (FPLO) basis set~\cite{Koepernik1999}. For the prediction of the hydrogen atom positions, we use a generalized gradient approximation (GGA) exchange correlation functional~\cite{Perdew1996}. We determine the Heisenberg Hamiltonian parameters from total energies where we correct for the strong electronic correlations on the Cu$^{2+}$ $3d$ orbitals by a DFT+U functional~\cite{Liechtenstein1995}. For the energy mapping approach, we classify all 4096 possible spin configurations for the twelve Cu$^{2+}$ ions in the unit cell after removing the symmetry. We find that the 364 distinct classical energies allow us to resolve twelve Heisenberg Hamiltonian parameters. We fix the Hund's rule coupling for Cu $3d$ as $J_{\rm H}=1$\,eV, in agreement with many previous studies~\cite{Iida2020,Chillal-2020}. We then perform the energy mapping approach for six values of the onsite interaction value $U$ (see Fig.~\ref{fig:Fig1}\,{\bf c}). We choose the appropriate value for spangolite by demanding that the Curie-Weiss temperature calculated as 
\begin{equation}\begin{split}
   \theta_{\rm CW}=\frac{1}{3}S(S+1) \big(&J_{1} + J_{2} + J_{3} + J_{4} + J_{5} + J_{15} + J_{16} \\&+ J_{17} + J_{18} + J_{19} + J_{21} + J_{23}\big)
\end{split}\end{equation}
agree with the experimental value of $\theta_{\rm CW}$. 
For this purpose, we reconsider the Curie-Weiss fit performed in Ref.~\onlinecite{Fennell2011}. We use a fitting {\it ansatz} developed recently by Pohle and Jaubert~\cite{Pohle2023} for spin liquids:
\begin{equation}
   \begin{split}
    &\chi T\big|^{\rm fit} =\frac{1+b_1\exp[c_1/T]}{a+b_2\exp[c_2/T]}\\
    &C=\frac{1+b_1}{a+b_2},\quad \theta_{\rm CW}=\frac{b_1c_1}{1+b_1}-\frac{b_2c_2}{a+b_2}
\end{split}\label{eq:ansatz}
\end{equation}
Note that this {\it ansatz} has recently allowed us to assign a Curie-Weiss temperature to nabokoite~\cite{Gonzalez2025}. The fit is excellent throughout the temperature range (see Supplementary Figure~1
) and allows us to revise the Curie-Weiss temperature of spangolite to $\theta_{\rm CW}=-29.4$\,K, which is somewhat smaller than the result of the traditional $\chi^{-1}$ fit of $\theta_{\rm CW}=-38$\,K. 

As a consistency check, we have evaluated the theoretical magnetic susceptibility that we obtained with the DFT Hamiltonian given in Table~\ref{tab:couplings} in the same way as the experimental susceptbility (see Supplementary Figure~2)
. The fit with the {\it ansatz} of Eq.~\eqref{eq:ansatz} yields Weiss constant $C$ and Curie-Weiss temperature $\theta_{\rm CW}$ that are in good agreement with experiment.

\subsection*{Tensor network ansatz}
The construction of the tensor network state proceeds by coarse-graining the two spins on the dominant anti\-ferro\-magnetic $J_1$ bonds in the maple-leaf lattice, which then maps it to a regular kagome lattice. Due to its corner-sharing triangles, a TN setup on the dual honeycomb lattice in the form of iPESS and iPESO is very well suited to capture the multipartite quantum correlations therein. The general TN setup is shown in Fig.~\ref{fig:Fig2}\,{\bf a}, where the stretched hexagons denote coarse-grained lattice sites, each represented by a single green tensor in {\bf b} and {\bf c}. In the chosen TN {\it ansatz}, those tensors reside on the links of the honeycomb lattice, so that additional three-index simplex tensors have to be introduced to connect them (gray tensors). Quantum states, represented by an iPESS, have only a single physical index per lattice site tensor, while thermal density matrices, represented by an iPESO, have two. This is shown in the figure as black, and both black and gray indices, respectively. A single unit cell consists of three lattice tensors (capturing six spins), together with two simplex tensors. Importantly, tensor networks in the thermodynamic limit offer the possibility of choosing arbitrary unit cells, that are repeated periodically to generate the 2D lattice. This gives us the possibility to determine the actual structure of the targeted states based on energy comparisons of different configurations~\cite{Schmoll-2023}. Each elementary unit cell of the honeycomb TN can be further coarse-grained into an effective tensor on a regular square lattice, as described in Supplementary Figure~4. It allows to combine six spins on the original maple-leaf lattice into a single tensor. This is important to compute accurate expectation values, for which we use a {corner transfer matrix renormalization group} (CTMRG)~\cite{Nishino1996,Nishino1997,Orus2009} procedure, as shown in Fig.~\ref{fig:Fig2}\,{\bf d}. The contraction of the infinite square lattice introduces an additional control parameter, the environment bond dimension $\chi_{\rm E}$. In our simulations we use system sizes of six, twelve and eighteen spins. Both ground and thermal state simulations are based on the very efficient simple update procedure~\cite{Jordan2008,Xie2014,Kshetrimayum2019,Schmoll2022}. It is an approximate, yet reliable algorithm frequently used in TN simulations. The main refinement parameters, i.e. the bond dimensions and the infinitesimal cooling step in the finite-temperature simulations, are carefully chosen to achieve sufficient accuracy and accessible temperature ranges while balancing the numerical effort. Details about the simple update for iPESS and iPESO, as well as the calculation of expectation values are presented in the Supplementary Note 3.

\section*{Data availability}

The datasets generated for and presented in this study are available at Ref.~\cite{SpangoliteData}.

\section*{Code Availability}

The DFT code used in this study is available from \texttt{https://www.fplo.de/}. 
The TN code developed during the present study is based on TensorKit.jl~\cite{TensorKit.jl} and available on GitHub~\cite{SpangoliteCode}.

\section*{Author contributions}

The project was conceived by HOJ and YI. HOJ performed the DFT and energy-mapping calculations. PS performed the tensor network calculations. PS, HOJ, and YI analysed and discussed the results, and wrote and reviewed the manuscript.

\section*{Competing interests}

The authors declare no competing interests.

\section*{Acknowledgments}

We acknowledge helpful discussions with Jan Naumann and Tobias M\"uller. 
This work has been funded by the Deutsche Forschungsgemeinschaft (DFG, German Research Foundation) under the project number 277101999 -- CRC 183 (project B01) and the BMBF (MUNIQC-Atoms, FermiQP). 
The authors would like to thank the HPC Service of ZEDAT, Freie Universität Berlin for computing time~\cite{Bennett2020}. H.~O.~J. acknowledges support through JSPS KAKENHI Grant No. 24H01668. 
The work of Y.I. was performed, in part, at the Aspen Center for Physics, which is supported by National Science Foundation Grant No.~PHY-2210452 and a grant from the Simons Foundation (1161654, Troyer). This research was supported in part by grant NSF PHY-2309135 to the Kavli Institute for Theoretical Physics (KITP).
Y.I. also acknowledges support from the ICTP through the Associates Programme and from the Simons Foundation through Grant No.~284558FY19, IIT Madras through the Institute of Eminence (IoE) program for establishing QuCenDiEM (Project No.~SP22231244CPETWOQCDHOC), and the International Centre for Theoretical Sciences (ICTS), Bengaluru, India during a visit for participating in the program Frustrated Metals and Insulators (Code No. ICTS/frumi2022/9). 
Y.I. further acknowledges the use of the computing resources at HPCE, IIT Madras. H.O.J. thanks IIT Madras for a visiting faculty fellow position during which work on this project was carried out.

\section*{References}

\begin{thebibliography}{64}%
\makeatletter
\providecommand \@ifxundefined [1]{%
 \@ifx{#1\undefined}
}%
\providecommand \@ifnum [1]{%
 \ifnum #1\expandafter \@firstoftwo
 \else \expandafter \@secondoftwo
 \fi
}%
\providecommand \@ifx [1]{%
 \ifx #1\expandafter \@firstoftwo
 \else \expandafter \@secondoftwo
 \fi
}%
\providecommand \natexlab [1]{#1}%
\providecommand \enquote  [1]{``#1''}%
\providecommand \bibnamefont  [1]{#1}%
\providecommand \bibfnamefont [1]{#1}%
\providecommand \citenamefont [1]{#1}%
\providecommand \href@noop [0]{\@secondoftwo}%
\providecommand \href [0]{\begingroup \@sanitize@url \@href}%
\providecommand \@href[1]{\@@startlink{#1}\@@href}%
\providecommand \@@href[1]{\endgroup#1\@@endlink}%
\providecommand \@sanitize@url [0]{\catcode `\\12\catcode `\$12\catcode
  `\&12\catcode `\#12\catcode `\^12\catcode `\_12\catcode `\%12\relax}%
\providecommand \@@startlink[1]{}%
\providecommand \@@endlink[0]{}%
\providecommand \url  [0]{\begingroup\@sanitize@url \@url }%
\providecommand \@url [1]{\endgroup\@href {#1}{\urlprefix }}%
\providecommand \urlprefix  [0]{URL }%
\providecommand \Eprint [0]{\href }%
\providecommand \doibase [0]{https://doi.org/}%
\providecommand \selectlanguage [0]{\@gobble}%
\providecommand \bibinfo  [0]{\@secondoftwo}%
\providecommand \bibfield  [0]{\@secondoftwo}%
\providecommand \translation [1]{[#1]}%
\providecommand \BibitemOpen [0]{}%
\providecommand \bibitemStop [0]{}%
\providecommand \bibitemNoStop [0]{.\EOS\space}%
\providecommand \EOS [0]{\spacefactor3000\relax}%
\providecommand \BibitemShut  [1]{\csname bibitem#1\endcsname}%
\let\auto@bib@innerbib\@empty
\bibitem [{\citenamefont {Huse}\ and\ \citenamefont
  {Rutenberg}(1992)}]{Huse-1992}%
  \BibitemOpen
  \bibfield  {author} {\bibinfo {author} {\bibfnamefont {D.~A.}\ \bibnamefont
  {Huse}}\ and\ \bibinfo {author} {\bibfnamefont {A.~D.}\ \bibnamefont
  {Rutenberg}},\ }\bibfield  {title} {\bibinfo {title} {{Classical
  antiferromagnets on the Kagom\'e lattice}},\ }\href
  {https://doi.org/10.1103/PhysRevB.45.7536} {\bibfield  {journal} {\bibinfo
  {journal} {Phys. Rev. B}\ }\textbf {\bibinfo {volume} {45}},\ \bibinfo
  {pages} {7536} (\bibinfo {year} {1992})}\BibitemShut {NoStop}%
\bibitem [{\citenamefont {Farnell}\ \emph {et~al.}(2011)\citenamefont
  {Farnell}, \citenamefont {Darradi}, \citenamefont {Schmidt},\ and\
  \citenamefont {Richter}}]{Farnell2011}%
  \BibitemOpen
  \bibfield  {author} {\bibinfo {author} {\bibfnamefont {D.~J.~J.}\
  \bibnamefont {Farnell}}, \bibinfo {author} {\bibfnamefont {R.}~\bibnamefont
  {Darradi}}, \bibinfo {author} {\bibfnamefont {R.}~\bibnamefont {Schmidt}},\
  and\ \bibinfo {author} {\bibfnamefont {J.}~\bibnamefont {Richter}},\
  }\bibfield  {title} {\bibinfo {title} {{Spin-half Heisenberg antiferromagnet
  on two archimedian lattices: From the bounce lattice to the maple-leaf
  lattice and beyond}},\ }\href {https://doi.org/10.1103/PhysRevB.84.104406}
  {\bibfield  {journal} {\bibinfo  {journal} {Phys. Rev. B}\ }\textbf {\bibinfo
  {volume} {84}},\ \bibinfo {pages} {104406} (\bibinfo {year}
  {2011})}\BibitemShut {NoStop}%
\bibitem [{\citenamefont {Farnell}\ \emph {et~al.}(2014)\citenamefont
  {Farnell}, \citenamefont {G\"otze}, \citenamefont {Richter}, \citenamefont
  {Bishop},\ and\ \citenamefont {Li}}]{Farnell-2014}%
  \BibitemOpen
  \bibfield  {author} {\bibinfo {author} {\bibfnamefont {D.~J.~J.}\
  \bibnamefont {Farnell}}, \bibinfo {author} {\bibfnamefont {O.}~\bibnamefont
  {G\"otze}}, \bibinfo {author} {\bibfnamefont {J.}~\bibnamefont {Richter}},
  \bibinfo {author} {\bibfnamefont {R.~F.}\ \bibnamefont {Bishop}},\ and\
  \bibinfo {author} {\bibfnamefont {P.~H.~Y.}\ \bibnamefont {Li}},\ }\bibfield
  {title} {\bibinfo {title} {{Quantum $s=\frac{1}{2}$ antiferromagnets on
  Archimedean lattices: The route from semiclassical magnetic order to
  nonmagnetic quantum states}},\ }\href
  {https://doi.org/10.1103/PhysRevB.89.184407} {\bibfield  {journal} {\bibinfo
  {journal} {Phys. Rev. B}\ }\textbf {\bibinfo {volume} {89}},\ \bibinfo
  {pages} {184407} (\bibinfo {year} {2014})}\BibitemShut {NoStop}%
\bibitem [{\citenamefont {Zheng}\ \emph {et~al.}(2014)\citenamefont {Zheng},
  \citenamefont {Zheng},\ and\ \citenamefont {Chen}}]{Zheng-2014}%
  \BibitemOpen
  \bibfield  {author} {\bibinfo {author} {\bibfnamefont {Y.-Z.}\ \bibnamefont
  {Zheng}}, \bibinfo {author} {\bibfnamefont {Z.}~\bibnamefont {Zheng}},\ and\
  \bibinfo {author} {\bibfnamefont {X.-M.}\ \bibnamefont {Chen}},\ }\bibfield
  {title} {\bibinfo {title} {{A symbol approach for classification of
  molecule-based magnetic materials exemplified by coordination polymers of
  metal carboxylates}},\ }\href
  {https://doi.org/https://doi.org/10.1016/j.ccr.2013.08.031} {\bibfield
  {journal} {\bibinfo  {journal} {Coord. Chem. Rev.}\ }\textbf {\bibinfo
  {volume} {258-259}},\ \bibinfo {pages} {1} (\bibinfo {year}
  {2014})}\BibitemShut {NoStop}%
\bibitem [{\citenamefont {Zheng}\ \emph {et~al.}(2007)\citenamefont {Zheng},
  \citenamefont {Tong}, \citenamefont {Xue}, \citenamefont {Zhang},
  \citenamefont {Chen}, \citenamefont {Grandjean},\ and\ \citenamefont
  {Long}}]{Zheng2007}%
  \BibitemOpen
  \bibfield  {author} {\bibinfo {author} {\bibfnamefont {Y.-Z.}\ \bibnamefont
  {Zheng}}, \bibinfo {author} {\bibfnamefont {M.-L.}\ \bibnamefont {Tong}},
  \bibinfo {author} {\bibfnamefont {W.}~\bibnamefont {Xue}}, \bibinfo {author}
  {\bibfnamefont {W.-X.}\ \bibnamefont {Zhang}}, \bibinfo {author}
  {\bibfnamefont {X.-M.}\ \bibnamefont {Chen}}, \bibinfo {author}
  {\bibfnamefont {F.}~\bibnamefont {Grandjean}},\ and\ \bibinfo {author}
  {\bibfnamefont {G.~J.}\ \bibnamefont {Long}},\ }\bibfield  {title} {\bibinfo
  {title} {A “star” antiferromagnet: A polymeric iron(iii) acetate that
  exhibits both spin frustration and long-range magnetic ordering},\ }\href
  {https://doi.org/10.1002/anie.200701954} {\bibfield  {journal} {\bibinfo
  {journal} {Angew. Chem. Int. Ed.}\ }\textbf {\bibinfo {volume} {46}},\
  \bibinfo {pages} {6076} (\bibinfo {year} {2007})}\BibitemShut {NoStop}%
\bibitem [{\citenamefont {Yamaguchi}\ \emph {et~al.}(2018)\citenamefont
  {Yamaguchi}, \citenamefont {Yoshizawa}, \citenamefont {Kida}, \citenamefont
  {Hagiwara}, \citenamefont {Matsuo}, \citenamefont {Kono}, \citenamefont
  {Sakakibara}, \citenamefont {Tamekuni}, \citenamefont {Miyagai},\ and\
  \citenamefont {Hosokoshi}}]{Yamaguchi2018}%
  \BibitemOpen
  \bibfield  {author} {\bibinfo {author} {\bibfnamefont {H.}~\bibnamefont
  {Yamaguchi}}, \bibinfo {author} {\bibfnamefont {D.}~\bibnamefont
  {Yoshizawa}}, \bibinfo {author} {\bibfnamefont {T.}~\bibnamefont {Kida}},
  \bibinfo {author} {\bibfnamefont {M.}~\bibnamefont {Hagiwara}}, \bibinfo
  {author} {\bibfnamefont {A.}~\bibnamefont {Matsuo}}, \bibinfo {author}
  {\bibfnamefont {Y.}~\bibnamefont {Kono}}, \bibinfo {author} {\bibfnamefont
  {T.}~\bibnamefont {Sakakibara}}, \bibinfo {author} {\bibfnamefont
  {Y.}~\bibnamefont {Tamekuni}}, \bibinfo {author} {\bibfnamefont
  {H.}~\bibnamefont {Miyagai}},\ and\ \bibinfo {author} {\bibfnamefont
  {Y.}~\bibnamefont {Hosokoshi}},\ }\bibfield  {title} {\bibinfo {title}
  {{Magnetic-field-induced Quantum Phase in S = 1/2 Frustrated Trellis
  Lattice}},\ }\href {https://doi.org/10.7566/JPSJ.87.043701} {\bibfield
  {journal} {\bibinfo  {journal} {J. Phys. Soc. Jpn.}\ }\textbf {\bibinfo
  {volume} {87}},\ \bibinfo {pages} {043701} (\bibinfo {year}
  {2018})}\BibitemShut {NoStop}%
\bibitem [{\citenamefont {Betts}(1995)}]{Betts-1995}%
  \BibitemOpen
  \bibfield  {author} {\bibinfo {author} {\bibfnamefont {D.}~\bibnamefont
  {Betts}},\ }\bibfield  {title} {\bibinfo {title} {{A new two-dimensional
  lattice of coordination number five}},\ }\href
  {http://hdl.handle.net/10222/35332} {\bibfield  {journal} {\bibinfo
  {journal} {Proc. N. S. Inst. Sci.}\ }\textbf {\bibinfo {volume} {40}},\
  \bibinfo {pages} {95} (\bibinfo {year} {1995})}\BibitemShut {NoStop}%
\bibitem [{\citenamefont {Norman}(2018)}]{Norman2018}%
  \BibitemOpen
  \bibfield  {author} {\bibinfo {author} {\bibfnamefont {M.}~\bibnamefont
  {Norman}},\ }\bibfield  {title} {\bibinfo {title} {Copper tellurium oxides
  – a playground for magnetism},\ }\href
  {https://doi.org/10.1016/j.jmmm.2017.11.006} {\bibfield  {journal} {\bibinfo
  {journal} {J. Mag. Mag. Mater.}\ }\textbf {\bibinfo {volume} {452}},\
  \bibinfo {pages} {507} (\bibinfo {year} {2018})}\BibitemShut {NoStop}%
\bibitem [{\citenamefont {Inosov}(2018)}]{Inosov2018}%
  \BibitemOpen
  \bibfield  {author} {\bibinfo {author} {\bibfnamefont {D.}~\bibnamefont
  {Inosov}},\ }\bibfield  {title} {\bibinfo {title} {Quantum magnetism in
  minerals},\ }\href {https://doi.org/10.1080/00018732.2018.1571986} {\bibfield
   {journal} {\bibinfo  {journal} {Adv. Phys.}\ }\textbf {\bibinfo {volume}
  {67}},\ \bibinfo {pages} {149} (\bibinfo {year} {2018})}\BibitemShut
  {NoStop}%
\bibitem [{\citenamefont {Mills}\ \emph {et~al.}(2014)\citenamefont {Mills},
  \citenamefont {Kampf}, \citenamefont {Christy}, \citenamefont {Housley},
  \citenamefont {Rossman}, \citenamefont {Reynolds},\ and\ \citenamefont
  {Marty}}]{Mills2014}%
  \BibitemOpen
  \bibfield  {author} {\bibinfo {author} {\bibfnamefont {S.~J.}\ \bibnamefont
  {Mills}}, \bibinfo {author} {\bibfnamefont {A.~R.}\ \bibnamefont {Kampf}},
  \bibinfo {author} {\bibfnamefont {A.~G.}\ \bibnamefont {Christy}}, \bibinfo
  {author} {\bibfnamefont {R.~M.}\ \bibnamefont {Housley}}, \bibinfo {author}
  {\bibfnamefont {G.~R.}\ \bibnamefont {Rossman}}, \bibinfo {author}
  {\bibfnamefont {R.~E.}\ \bibnamefont {Reynolds}},\ and\ \bibinfo {author}
  {\bibfnamefont {J.}~\bibnamefont {Marty}},\ }\bibfield  {title} {\bibinfo
  {title} {Bluebellite and mojaveite, two new minerals from the central
  {M}ojave {D}esert, {C}alifornia, {USA}},\ }\href
  {https://doi.org/10.1180/minmag.2014.078.5.15} {\bibfield  {journal}
  {\bibinfo  {journal} {Mineralog. Mag.}\ }\textbf {\bibinfo {volume} {78}},\
  \bibinfo {pages} {1325–1340} (\bibinfo {year} {2014})}\BibitemShut
  {NoStop}%
\bibitem [{\citenamefont {Kampf}\ \emph {et~al.}(2013)\citenamefont {Kampf},
  \citenamefont {Mills}, \citenamefont {Housley},\ and\ \citenamefont
  {Marty}}]{Kampf2013}%
  \BibitemOpen
  \bibfield  {author} {\bibinfo {author} {\bibfnamefont {A.~R.}\ \bibnamefont
  {Kampf}}, \bibinfo {author} {\bibfnamefont {S.~J.}\ \bibnamefont {Mills}},
  \bibinfo {author} {\bibfnamefont {R.~M.}\ \bibnamefont {Housley}},\ and\
  \bibinfo {author} {\bibfnamefont {J.}~\bibnamefont {Marty}},\ }\bibfield
  {title} {\bibinfo {title} {Lead-tellurium oxysalts from {O}tto {M}ountain
  near {B}aker, {C}alifornia: {VIII}. fuettererite, \ce{Pb3Cu^{2+}6
  Te^{6+}O6(OH)7Cl5}, a new mineral with double spangolite-type sheets},\
  }\href {https://doi.org/10.2138/am.2013.4218} {\bibfield  {journal} {\bibinfo
   {journal} {Am. Mineralog.}\ }\textbf {\bibinfo {volume} {98}},\ \bibinfo
  {pages} {506} (\bibinfo {year} {2013})}\BibitemShut {NoStop}%
\bibitem [{\citenamefont {Olmi}\ \emph {et~al.}(1995)\citenamefont {Olmi},
  \citenamefont {Sabelli},\ and\ \citenamefont {Trosti-Ferroni}}]{Olmi1995}%
  \BibitemOpen
  \bibfield  {author} {\bibinfo {author} {\bibfnamefont {F.}~\bibnamefont
  {Olmi}}, \bibinfo {author} {\bibfnamefont {C.}~\bibnamefont {Sabelli}},\ and\
  \bibinfo {author} {\bibfnamefont {R.}~\bibnamefont {Trosti-Ferroni}},\
  }\bibfield  {title} {\bibinfo {title} {The crystal structure of sabelliite},\
  }\href {https://doi.org/10.1127/ejm/7/6/1331} {\bibfield  {journal} {\bibinfo
   {journal} {Eur. J. Mineral.}\ }\textbf {\bibinfo {volume} {7}},\ \bibinfo
  {pages} {1331} (\bibinfo {year} {1995})}\BibitemShut {NoStop}%
\bibitem [{\citenamefont {Penfield}(1890)}]{Penfield1890}%
  \BibitemOpen
  \bibfield  {author} {\bibinfo {author} {\bibfnamefont {S.~L.}\ \bibnamefont
  {Penfield}},\ }\bibfield  {title} {\bibinfo {title} {On spangolite, a new
  copper mineral},\ }\href {https://doi.org/10.2475/ajs.s3-39.233.370}
  {\bibfield  {journal} {\bibinfo  {journal} {Am. J. Sci.}\ }\textbf {\bibinfo
  {volume} {39}},\ \bibinfo {pages} {370} (\bibinfo {year} {1890})}\BibitemShut
  {NoStop}%
\bibitem [{\citenamefont {Miers}(1893)}]{Miers-1893}%
  \BibitemOpen
  \bibfield  {author} {\bibinfo {author} {\bibfnamefont {H.~A.}\ \bibnamefont
  {Miers}},\ }\bibfield  {title} {\bibinfo {title} {{Spangolite, a Remarkable
  Cornish Mineral}},\ }\href {https://doi.org/10.1038/048426b0} {\bibfield
  {journal} {\bibinfo  {journal} {Nature}\ }\textbf {\bibinfo {volume} {48}},\
  \bibinfo {pages} {426} (\bibinfo {year} {1893})}\BibitemShut {NoStop}%
\bibitem [{\citenamefont {Frondel}(1949)}]{Frondel-1949}%
  \BibitemOpen
  \bibfield  {author} {\bibinfo {author} {\bibfnamefont {C.}~\bibnamefont
  {Frondel}},\ }\bibfield  {title} {\bibinfo {title} {{Crystallography of
  spangolite}},\ }\href {http://www.minsocam.org/ammin/AM34/AM34_181.pdf}
  {\bibfield  {journal} {\bibinfo  {journal} {Am. Mineral.}\ }\textbf {\bibinfo
  {volume} {34}},\ \bibinfo {pages} {181} (\bibinfo {year} {1949})}\BibitemShut
  {NoStop}%
\bibitem [{\citenamefont {Hawthorne}\ \emph {et~al.}(1993)\citenamefont
  {Hawthorne}, \citenamefont {Kimata},\ and\ \citenamefont
  {Eby}}]{Hawthorne1993}%
  \BibitemOpen
  \bibfield  {author} {\bibinfo {author} {\bibfnamefont {F.~C.}\ \bibnamefont
  {Hawthorne}}, \bibinfo {author} {\bibfnamefont {M.}~\bibnamefont {Kimata}},\
  and\ \bibinfo {author} {\bibfnamefont {R.~K.}\ \bibnamefont {Eby}},\
  }\bibfield  {title} {\bibinfo {title} {{The crystal structure of spangolite,
  a complex copper sulfate sheet mineral}},\ }\href
  {http://www.minsocam.org/ammin/AM78/AM78_649.pdf} {\bibfield  {journal}
  {\bibinfo  {journal} {Am. Mineral.}\ }\textbf {\bibinfo {volume} {78}},\
  \bibinfo {pages} {649} (\bibinfo {year} {1993})}\BibitemShut {NoStop}%
\bibitem [{\citenamefont {Haraguchi}\ \emph {et~al.}(2018)\citenamefont
  {Haraguchi}, \citenamefont {Matsuo}, \citenamefont {Kindo},\ and\
  \citenamefont {Hiroi}}]{Haraguchi2018}%
  \BibitemOpen
  \bibfield  {author} {\bibinfo {author} {\bibfnamefont {Y.}~\bibnamefont
  {Haraguchi}}, \bibinfo {author} {\bibfnamefont {A.}~\bibnamefont {Matsuo}},
  \bibinfo {author} {\bibfnamefont {K.}~\bibnamefont {Kindo}},\ and\ \bibinfo
  {author} {\bibfnamefont {Z.}~\bibnamefont {Hiroi}},\ }\bibfield  {title}
  {\bibinfo {title} {{Frustrated magnetism of the maple-leaf-lattice
  antiferromagnet MgMn$_3$O$_7$$\cdot$3H$_{2}$O}},\ }\href
  {https://doi.org/10.1103/PhysRevB.98.064412} {\bibfield  {journal} {\bibinfo
  {journal} {Phys. Rev. B}\ }\textbf {\bibinfo {volume} {98}},\ \bibinfo
  {pages} {064412} (\bibinfo {year} {2018})}\BibitemShut {NoStop}%
\bibitem [{\citenamefont {Venkatesh}\ \emph {et~al.}(2020)\citenamefont
  {Venkatesh}, \citenamefont {Bandyopadhyay}, \citenamefont {Midya},
  \citenamefont {Mahalingam}, \citenamefont {Ganesan},\ and\ \citenamefont
  {Mandal}}]{Venkatesh2020}%
  \BibitemOpen
  \bibfield  {author} {\bibinfo {author} {\bibfnamefont {C.}~\bibnamefont
  {Venkatesh}}, \bibinfo {author} {\bibfnamefont {B.}~\bibnamefont
  {Bandyopadhyay}}, \bibinfo {author} {\bibfnamefont {A.}~\bibnamefont
  {Midya}}, \bibinfo {author} {\bibfnamefont {K.}~\bibnamefont {Mahalingam}},
  \bibinfo {author} {\bibfnamefont {V.}~\bibnamefont {Ganesan}},\ and\ \bibinfo
  {author} {\bibfnamefont {P.}~\bibnamefont {Mandal}},\ }\bibfield  {title}
  {\bibinfo {title} {{Magnetic properties of the one-dimensional
  $S=\frac{3}{2}$ Heisenberg antiferromagnetic spin-chain compound
  \ce{Na2Mn3O7}}},\ }\href {https://doi.org/10.1103/PhysRevB.101.184429}
  {\bibfield  {journal} {\bibinfo  {journal} {Phys. Rev. B}\ }\textbf {\bibinfo
  {volume} {101}},\ \bibinfo {pages} {184429} (\bibinfo {year}
  {2020})}\BibitemShut {NoStop}%
\bibitem [{\citenamefont {Saha}\ \emph {et~al.}(2023)\citenamefont {Saha},
  \citenamefont {Bera}, \citenamefont {Yusuf},\ and\ \citenamefont
  {Hoser}}]{Saha-2023}%
  \BibitemOpen
  \bibfield  {author} {\bibinfo {author} {\bibfnamefont {B.}~\bibnamefont
  {Saha}}, \bibinfo {author} {\bibfnamefont {A.~K.}\ \bibnamefont {Bera}},
  \bibinfo {author} {\bibfnamefont {S.~M.}\ \bibnamefont {Yusuf}},\ and\
  \bibinfo {author} {\bibfnamefont {A.}~\bibnamefont {Hoser}},\ }\bibfield
  {title} {\bibinfo {title} {{Two-dimensional short-range spin-spin
  correlations in the layered spin-$\frac{3}{2}$ maple leaf lattice
  antiferromagnet ${\mathrm{Na}}_{2}{\mathrm{Mn}}_{3}{\mathrm{O}}_{7}$ with
  crystal stacking disorder}},\ }\href
  {https://doi.org/10.1103/PhysRevB.107.064419} {\bibfield  {journal} {\bibinfo
   {journal} {Phys. Rev. B}\ }\textbf {\bibinfo {volume} {107}},\ \bibinfo
  {pages} {064419} (\bibinfo {year} {2023})}\BibitemShut {NoStop}%
\bibitem [{\citenamefont {Fennell}\ \emph {et~al.}(2011)\citenamefont
  {Fennell}, \citenamefont {Piatek}, \citenamefont {Stephenson}, \citenamefont
  {Nilsen},\ and\ \citenamefont {Rønnow}}]{Fennell2011}%
  \BibitemOpen
  \bibfield  {author} {\bibinfo {author} {\bibfnamefont {T.}~\bibnamefont
  {Fennell}}, \bibinfo {author} {\bibfnamefont {J.~O.}\ \bibnamefont {Piatek}},
  \bibinfo {author} {\bibfnamefont {R.~A.}\ \bibnamefont {Stephenson}},
  \bibinfo {author} {\bibfnamefont {G.~J.}\ \bibnamefont {Nilsen}},\ and\
  \bibinfo {author} {\bibfnamefont {H.~M.}\ \bibnamefont {Rønnow}},\
  }\bibfield  {title} {\bibinfo {title} {Spangolite: an s = 1/2 maple leaf
  lattice antiferromagnet?},\ }\href
  {https://doi.org/10.1088/0953-8984/23/16/164201} {\bibfield  {journal}
  {\bibinfo  {journal} {J. Phys.: Condens. Matter}\ }\textbf {\bibinfo {volume}
  {23}},\ \bibinfo {pages} {164201} (\bibinfo {year} {2011})}\BibitemShut
  {NoStop}%
\bibitem [{\citenamefont {Xie}\ \emph {et~al.}(2014)\citenamefont {Xie},
  \citenamefont {Chen}, \citenamefont {Yu}, \citenamefont {Kong}, \citenamefont
  {Normand},\ and\ \citenamefont {Xiang}}]{Xie2014}%
  \BibitemOpen
  \bibfield  {author} {\bibinfo {author} {\bibfnamefont {Z.~Y.}\ \bibnamefont
  {Xie}}, \bibinfo {author} {\bibfnamefont {J.}~\bibnamefont {Chen}}, \bibinfo
  {author} {\bibfnamefont {J.~F.}\ \bibnamefont {Yu}}, \bibinfo {author}
  {\bibfnamefont {X.}~\bibnamefont {Kong}}, \bibinfo {author} {\bibfnamefont
  {B.}~\bibnamefont {Normand}},\ and\ \bibinfo {author} {\bibfnamefont
  {T.}~\bibnamefont {Xiang}},\ }\bibfield  {title} {\bibinfo {title} {Tensor
  renormalization of quantum many-body systems using projected entangled
  simplex states},\ }\href {https://doi.org/10.1103/PhysRevX.4.011025}
  {\bibfield  {journal} {\bibinfo  {journal} {Phys. Rev. X}\ }\textbf {\bibinfo
  {volume} {4}},\ \bibinfo {pages} {011025} (\bibinfo {year}
  {2014})}\BibitemShut {NoStop}%
\bibitem [{\citenamefont {Schmoll}\ \emph {et~al.}(2024)\citenamefont
  {Schmoll}, \citenamefont {Balz}, \citenamefont {Lake}, \citenamefont
  {Eisert},\ and\ \citenamefont {Kshetrimayum}}]{Schmoll2022}%
  \BibitemOpen
  \bibfield  {author} {\bibinfo {author} {\bibfnamefont {P.}~\bibnamefont
  {Schmoll}}, \bibinfo {author} {\bibfnamefont {C.}~\bibnamefont {Balz}},
  \bibinfo {author} {\bibfnamefont {B.}~\bibnamefont {Lake}}, \bibinfo {author}
  {\bibfnamefont {J.}~\bibnamefont {Eisert}},\ and\ \bibinfo {author}
  {\bibfnamefont {A.}~\bibnamefont {Kshetrimayum}},\ }\bibfield  {title}
  {\bibinfo {title} {{Finite temperature tensor network algorithm for
  frustrated two-dimensional quantum materials}},\ }\href
  {https://doi.org/10.1103/PhysRevB.109.235119} {\bibfield  {journal} {\bibinfo
   {journal} {Phys. Rev. B}\ }\textbf {\bibinfo {volume} {109}},\ \bibinfo
  {pages} {235119} (\bibinfo {year} {2024})}\BibitemShut {NoStop}%
\bibitem [{\citenamefont {Ghosh}\ \emph {et~al.}(2022)\citenamefont {Ghosh},
  \citenamefont {M\"uller},\ and\ \citenamefont {Thomale}}]{Ghosh-2022}%
  \BibitemOpen
  \bibfield  {author} {\bibinfo {author} {\bibfnamefont {P.}~\bibnamefont
  {Ghosh}}, \bibinfo {author} {\bibfnamefont {T.}~\bibnamefont {M\"uller}},\
  and\ \bibinfo {author} {\bibfnamefont {R.}~\bibnamefont {Thomale}},\
  }\bibfield  {title} {\bibinfo {title} {{Another exact ground state of a
  two-dimensional quantum antiferromagnet}},\ }\href
  {https://doi.org/10.1103/PhysRevB.105.L180412} {\bibfield  {journal}
  {\bibinfo  {journal} {Phys. Rev. B}\ }\textbf {\bibinfo {volume} {105}},\
  \bibinfo {pages} {L180412} (\bibinfo {year} {2022})}\BibitemShut {NoStop}%
\bibitem [{\citenamefont {Makuta}\ and\ \citenamefont
  {Hotta}(2021)}]{Makuta2021}%
  \BibitemOpen
  \bibfield  {author} {\bibinfo {author} {\bibfnamefont {R.}~\bibnamefont
  {Makuta}}\ and\ \bibinfo {author} {\bibfnamefont {C.}~\bibnamefont {Hotta}},\
  }\bibfield  {title} {\bibinfo {title} {Dimensional reduction in quantum
  spin-$\frac{1}{2}$ system on a $\frac{1}{7}$-depleted triangular lattice},\
  }\href {https://doi.org/10.1103/PhysRevB.104.224415} {\bibfield  {journal}
  {\bibinfo  {journal} {Phys. Rev. B}\ }\textbf {\bibinfo {volume} {104}},\
  \bibinfo {pages} {224415} (\bibinfo {year} {2021})}\BibitemShut {NoStop}%
\bibitem [{\citenamefont {Ghosh}\ \emph {et~al.}(2024)\citenamefont {Ghosh},
  \citenamefont {M\"uller}, \citenamefont {Iqbal}, \citenamefont {Thomale},\
  and\ \citenamefont {Jeschke}}]{Ghosh-2023_bluebellite}%
  \BibitemOpen
  \bibfield  {author} {\bibinfo {author} {\bibfnamefont {P.}~\bibnamefont
  {Ghosh}}, \bibinfo {author} {\bibfnamefont {T.}~\bibnamefont {M\"uller}},
  \bibinfo {author} {\bibfnamefont {Y.}~\bibnamefont {Iqbal}}, \bibinfo
  {author} {\bibfnamefont {R.}~\bibnamefont {Thomale}},\ and\ \bibinfo {author}
  {\bibfnamefont {H.~O.}\ \bibnamefont {Jeschke}},\ }\bibfield  {title}
  {\bibinfo {title} {{Effective spin-1 breathing kagome Hamiltonian induced by
  the exchange hierarchy in the maple leaf mineral bluebellite}},\ }\href
  {https://doi.org/10.1103/PhysRevB.110.094406} {\bibfield  {journal} {\bibinfo
   {journal} {Phys. Rev. B}\ }\textbf {\bibinfo {volume} {110}},\ \bibinfo
  {pages} {094406} (\bibinfo {year} {2024})}\BibitemShut {NoStop}%
\bibitem [{\citenamefont {Gresista}\ \emph {et~al.}(2023)\citenamefont
  {Gresista}, \citenamefont {Hickey}, \citenamefont {Trebst},\ and\
  \citenamefont {Iqbal}}]{Gresista-2023}%
  \BibitemOpen
  \bibfield  {author} {\bibinfo {author} {\bibfnamefont {L.}~\bibnamefont
  {Gresista}}, \bibinfo {author} {\bibfnamefont {C.}~\bibnamefont {Hickey}},
  \bibinfo {author} {\bibfnamefont {S.}~\bibnamefont {Trebst}},\ and\ \bibinfo
  {author} {\bibfnamefont {Y.}~\bibnamefont {Iqbal}},\ }\bibfield  {title}
  {\bibinfo {title} {{Candidate quantum disordered intermediate phase in the
  Heisenberg antiferromagnet on the maple-leaf lattice}},\ }\href
  {https://doi.org/10.1103/PhysRevB.108.L241116} {\bibfield  {journal}
  {\bibinfo  {journal} {Phys. Rev. B}\ }\textbf {\bibinfo {volume} {108}},\
  \bibinfo {pages} {L241116} (\bibinfo {year} {2023})}\BibitemShut {NoStop}%
\bibitem [{\citenamefont {Iida}\ \emph {et~al.}(2020)\citenamefont {Iida},
  \citenamefont {Yoshida}, \citenamefont {Nakao}, \citenamefont {Jeschke},
  \citenamefont {Iqbal}, \citenamefont {Nakajima}, \citenamefont
  {Ohira-Kawamura}, \citenamefont {Munakata}, \citenamefont {Inamura},
  \citenamefont {Murai}, \citenamefont {Ishikado}, \citenamefont {Kumai},
  \citenamefont {Okada}, \citenamefont {Oda}, \citenamefont {Kakurai},\ and\
  \citenamefont {Matsuda}}]{Iida2020}%
  \BibitemOpen
  \bibfield  {author} {\bibinfo {author} {\bibfnamefont {K.}~\bibnamefont
  {Iida}}, \bibinfo {author} {\bibfnamefont {H.~K.}\ \bibnamefont {Yoshida}},
  \bibinfo {author} {\bibfnamefont {A.}~\bibnamefont {Nakao}}, \bibinfo
  {author} {\bibfnamefont {H.~O.}\ \bibnamefont {Jeschke}}, \bibinfo {author}
  {\bibfnamefont {Y.}~\bibnamefont {Iqbal}}, \bibinfo {author} {\bibfnamefont
  {K.}~\bibnamefont {Nakajima}}, \bibinfo {author} {\bibfnamefont
  {S.}~\bibnamefont {Ohira-Kawamura}}, \bibinfo {author} {\bibfnamefont
  {K.}~\bibnamefont {Munakata}}, \bibinfo {author} {\bibfnamefont
  {Y.}~\bibnamefont {Inamura}}, \bibinfo {author} {\bibfnamefont
  {N.}~\bibnamefont {Murai}}, \bibinfo {author} {\bibfnamefont
  {M.}~\bibnamefont {Ishikado}}, \bibinfo {author} {\bibfnamefont
  {R.}~\bibnamefont {Kumai}}, \bibinfo {author} {\bibfnamefont
  {T.}~\bibnamefont {Okada}}, \bibinfo {author} {\bibfnamefont
  {M.}~\bibnamefont {Oda}}, \bibinfo {author} {\bibfnamefont {K.}~\bibnamefont
  {Kakurai}},\ and\ \bibinfo {author} {\bibfnamefont {M.}~\bibnamefont
  {Matsuda}},\ }\bibfield  {title} {\bibinfo {title} {$q=0$ long-range magnetic
  order in centennialite \ce{CaCu3(OD)6Cl2\cdot {0.6}D2O}: A spin-$\frac{1}{2}$
  perfect kagome antiferromagnet with ${J}_{1}$-${J}_{2}$-${J}_{d}$},\ }\href
  {https://doi.org/10.1103/PhysRevB.101.220408} {\bibfield  {journal} {\bibinfo
   {journal} {Phys. Rev. B}\ }\textbf {\bibinfo {volume} {101}},\ \bibinfo
  {pages} {220408} (\bibinfo {year} {2020})}\BibitemShut {NoStop}%
\bibitem [{\citenamefont {Yamamoto}\ \emph {et~al.}(2021)\citenamefont
  {Yamamoto}, \citenamefont {Sakakura}, \citenamefont {Jeschke}, \citenamefont
  {Kabeya}, \citenamefont {Hayashi}, \citenamefont {Ishikawa}, \citenamefont
  {Fujii}, \citenamefont {Kishimoto}, \citenamefont {Sagayama}, \citenamefont
  {Shigematsu}, \citenamefont {Azuma}, \citenamefont {Ochiai}, \citenamefont
  {Noda},\ and\ \citenamefont {Kimura}}]{Yamamoto2021}%
  \BibitemOpen
  \bibfield  {author} {\bibinfo {author} {\bibfnamefont {H.}~\bibnamefont
  {Yamamoto}}, \bibinfo {author} {\bibfnamefont {T.}~\bibnamefont {Sakakura}},
  \bibinfo {author} {\bibfnamefont {H.~O.}\ \bibnamefont {Jeschke}}, \bibinfo
  {author} {\bibfnamefont {N.}~\bibnamefont {Kabeya}}, \bibinfo {author}
  {\bibfnamefont {K.}~\bibnamefont {Hayashi}}, \bibinfo {author} {\bibfnamefont
  {Y.}~\bibnamefont {Ishikawa}}, \bibinfo {author} {\bibfnamefont
  {Y.}~\bibnamefont {Fujii}}, \bibinfo {author} {\bibfnamefont
  {S.}~\bibnamefont {Kishimoto}}, \bibinfo {author} {\bibfnamefont
  {H.}~\bibnamefont {Sagayama}}, \bibinfo {author} {\bibfnamefont
  {K.}~\bibnamefont {Shigematsu}}, \bibinfo {author} {\bibfnamefont
  {M.}~\bibnamefont {Azuma}}, \bibinfo {author} {\bibfnamefont
  {A.}~\bibnamefont {Ochiai}}, \bibinfo {author} {\bibfnamefont
  {Y.}~\bibnamefont {Noda}},\ and\ \bibinfo {author} {\bibfnamefont
  {H.}~\bibnamefont {Kimura}},\ }\bibfield  {title} {\bibinfo {title} {{Quantum
  spin fluctuations and hydrogen bond network in the antiferromagnetic natural
  mineral henmilite}},\ }\href
  {https://doi.org/10.1103/PhysRevMaterials.5.104405} {\bibfield  {journal}
  {\bibinfo  {journal} {Phys. Rev. Mater.}\ }\textbf {\bibinfo {volume} {5}},\
  \bibinfo {pages} {104405} (\bibinfo {year} {2021})}\BibitemShut {NoStop}%
\bibitem [{\citenamefont {Fujihala}\ \emph {et~al.}(2022)\citenamefont
  {Fujihala}, \citenamefont {Jeschke}, \citenamefont {Morita}, \citenamefont
  {Kuwai}, \citenamefont {Koda}, \citenamefont {Okabe}, \citenamefont {Matsuo},
  \citenamefont {Kindo},\ and\ \citenamefont {Mitsuda}}]{Fujihara2022}%
  \BibitemOpen
  \bibfield  {author} {\bibinfo {author} {\bibfnamefont {M.}~\bibnamefont
  {Fujihala}}, \bibinfo {author} {\bibfnamefont {H.~O.}\ \bibnamefont
  {Jeschke}}, \bibinfo {author} {\bibfnamefont {K.}~\bibnamefont {Morita}},
  \bibinfo {author} {\bibfnamefont {T.}~\bibnamefont {Kuwai}}, \bibinfo
  {author} {\bibfnamefont {A.}~\bibnamefont {Koda}}, \bibinfo {author}
  {\bibfnamefont {H.}~\bibnamefont {Okabe}}, \bibinfo {author} {\bibfnamefont
  {A.}~\bibnamefont {Matsuo}}, \bibinfo {author} {\bibfnamefont
  {K.}~\bibnamefont {Kindo}},\ and\ \bibinfo {author} {\bibfnamefont
  {S.}~\bibnamefont {Mitsuda}},\ }\bibfield  {title} {\bibinfo {title}
  {{Birchite \ce{Cd2Cu2(PO4)2SO4}$\cdot 5{\rm H}_2$O as a model
  antiferromagnetic spin-1/2 Heisenberg ${J}_{1}\text{\ensuremath{-}}{J}_{2}$
  chain}},\ }\href {https://doi.org/10.1103/PhysRevMaterials.6.114408}
  {\bibfield  {journal} {\bibinfo  {journal} {Phys. Rev. Mater.}\ }\textbf
  {\bibinfo {volume} {6}},\ \bibinfo {pages} {114408} (\bibinfo {year}
  {2022})}\BibitemShut {NoStop}%
\bibitem [{\citenamefont {Iqbal}\ \emph {et~al.}(2015)\citenamefont {Iqbal},
  \citenamefont {Jeschke}, \citenamefont {Reuther}, \citenamefont
  {Valent\'{\i}}, \citenamefont {Mazin}, \citenamefont {Greiter},\ and\
  \citenamefont {Thomale}}]{Iqbal-2015}%
  \BibitemOpen
  \bibfield  {author} {\bibinfo {author} {\bibfnamefont {Y.}~\bibnamefont
  {Iqbal}}, \bibinfo {author} {\bibfnamefont {H.~O.}\ \bibnamefont {Jeschke}},
  \bibinfo {author} {\bibfnamefont {J.}~\bibnamefont {Reuther}}, \bibinfo
  {author} {\bibfnamefont {R.}~\bibnamefont {Valent\'{\i}}}, \bibinfo {author}
  {\bibfnamefont {I.~I.}\ \bibnamefont {Mazin}}, \bibinfo {author}
  {\bibfnamefont {M.}~\bibnamefont {Greiter}},\ and\ \bibinfo {author}
  {\bibfnamefont {R.}~\bibnamefont {Thomale}},\ }\bibfield  {title} {\bibinfo
  {title} {{Paramagnetism in the kagome compounds
  $(\mathrm{Zn},\mathrm{Mg},\mathrm{Cd}){\mathrm{Cu}}_{3}{(\mathrm{OH})}_{6}{\mathrm{Cl}}_{2}$}},\
  }\href {https://doi.org/10.1103/PhysRevB.92.220404} {\bibfield  {journal}
  {\bibinfo  {journal} {Phys. Rev. B}\ }\textbf {\bibinfo {volume} {92}},\
  \bibinfo {pages} {220404} (\bibinfo {year} {2015})}\BibitemShut {NoStop}%
\bibitem [{\citenamefont {Powell}\ \emph {et~al.}(1994)\citenamefont {Powell},
  \citenamefont {Thomas}, \citenamefont {Williams}, \citenamefont {Birch},\
  and\ \citenamefont {Plimer}}]{Powell1994}%
  \BibitemOpen
  \bibfield  {author} {\bibinfo {author} {\bibfnamefont {D.~W.}\ \bibnamefont
  {Powell}}, \bibinfo {author} {\bibfnamefont {R.~G.}\ \bibnamefont {Thomas}},
  \bibinfo {author} {\bibfnamefont {P.~A.}\ \bibnamefont {Williams}}, \bibinfo
  {author} {\bibfnamefont {W.~D.}\ \bibnamefont {Birch}},\ and\ \bibinfo
  {author} {\bibfnamefont {I.~R.}\ \bibnamefont {Plimer}},\ }\bibfield  {title}
  {\bibinfo {title} {Choloalite: synthesis and revised chemical formula},\
  }\href {https://doi.org/10.1180/minmag.1994.058.392.20} {\bibfield  {journal}
  {\bibinfo  {journal} {Mineralog. Mag.}\ }\textbf {\bibinfo {volume} {58}},\
  \bibinfo {pages} {505} (\bibinfo {year} {1994})}\BibitemShut {NoStop}%
\bibitem [{\citenamefont {Chillal}\ \emph {et~al.}(2020)\citenamefont
  {Chillal}, \citenamefont {Iqbal}, \citenamefont {Jeschke}, \citenamefont
  {Rodriguez-Rivera}, \citenamefont {Bewley}, \citenamefont {Manuel},
  \citenamefont {Khalyavin}, \citenamefont {Steffens}, \citenamefont {Thomale},
  \citenamefont {Islam}, \citenamefont {Reuther},\ and\ \citenamefont
  {Lake}}]{Chillal-2020}%
  \BibitemOpen
  \bibfield  {author} {\bibinfo {author} {\bibfnamefont {S.}~\bibnamefont
  {Chillal}}, \bibinfo {author} {\bibfnamefont {Y.}~\bibnamefont {Iqbal}},
  \bibinfo {author} {\bibfnamefont {H.~O.}\ \bibnamefont {Jeschke}}, \bibinfo
  {author} {\bibfnamefont {J.~A.}\ \bibnamefont {Rodriguez-Rivera}}, \bibinfo
  {author} {\bibfnamefont {R.}~\bibnamefont {Bewley}}, \bibinfo {author}
  {\bibfnamefont {P.}~\bibnamefont {Manuel}}, \bibinfo {author} {\bibfnamefont
  {D.}~\bibnamefont {Khalyavin}}, \bibinfo {author} {\bibfnamefont
  {P.}~\bibnamefont {Steffens}}, \bibinfo {author} {\bibfnamefont
  {R.}~\bibnamefont {Thomale}}, \bibinfo {author} {\bibfnamefont {A.~T. M.~N.}\
  \bibnamefont {Islam}}, \bibinfo {author} {\bibfnamefont {J.}~\bibnamefont
  {Reuther}},\ and\ \bibinfo {author} {\bibfnamefont {B.}~\bibnamefont
  {Lake}},\ }\bibfield  {title} {\bibinfo {title} {{Evidence for a
  three-dimensional quantum spin liquid in PbCuTe$_2$O$_6$}},\ }\href
  {https://doi.org/10.1038/s41467-020-15594-1} {\bibfield  {journal} {\bibinfo
  {journal} {Nat. Commun.}\ }\textbf {\bibinfo {volume} {11}},\ \bibinfo
  {pages} {2348} (\bibinfo {year} {2020})}\BibitemShut {NoStop}%
\bibitem [{\citenamefont {Koepernik}\ and\ \citenamefont
  {Eschrig}(1999)}]{Koepernik1999}%
  \BibitemOpen
  \bibfield  {author} {\bibinfo {author} {\bibfnamefont {K.}~\bibnamefont
  {Koepernik}}\ and\ \bibinfo {author} {\bibfnamefont {H.}~\bibnamefont
  {Eschrig}},\ }\bibfield  {title} {\bibinfo {title} {Full-potential
  nonorthogonal local-orbital minimum-basis band-structure scheme},\ }\href
  {https://doi.org/10.1103/PhysRevB.59.1743} {\bibfield  {journal} {\bibinfo
  {journal} {Phys. Rev. B}\ }\textbf {\bibinfo {volume} {59}},\ \bibinfo
  {pages} {1743} (\bibinfo {year} {1999})}\BibitemShut {NoStop}%
\bibitem [{\citenamefont {Perdew}\ \emph {et~al.}(1996)\citenamefont {Perdew},
  \citenamefont {Burke},\ and\ \citenamefont {Ernzerhof}}]{Perdew1996}%
  \BibitemOpen
  \bibfield  {author} {\bibinfo {author} {\bibfnamefont {J.~P.}\ \bibnamefont
  {Perdew}}, \bibinfo {author} {\bibfnamefont {K.}~\bibnamefont {Burke}},\ and\
  \bibinfo {author} {\bibfnamefont {M.}~\bibnamefont {Ernzerhof}},\ }\bibfield
  {title} {\bibinfo {title} {Generalized gradient approximation made simple},\
  }\href {https://doi.org/10.1103/PhysRevLett.77.3865} {\bibfield  {journal}
  {\bibinfo  {journal} {Phys. Rev. Lett.}\ }\textbf {\bibinfo {volume} {77}},\
  \bibinfo {pages} {3865} (\bibinfo {year} {1996})}\BibitemShut {NoStop}%
\bibitem [{\citenamefont {Haraguchi}\ \emph {et~al.}(2021)\citenamefont
  {Haraguchi}, \citenamefont {Matsuo}, \citenamefont {Kindo},\ and\
  \citenamefont {Hiroi}}]{Haraguchi2021}%
  \BibitemOpen
  \bibfield  {author} {\bibinfo {author} {\bibfnamefont {Y.}~\bibnamefont
  {Haraguchi}}, \bibinfo {author} {\bibfnamefont {A.}~\bibnamefont {Matsuo}},
  \bibinfo {author} {\bibfnamefont {K.}~\bibnamefont {Kindo}},\ and\ \bibinfo
  {author} {\bibfnamefont {Z.}~\bibnamefont {Hiroi}},\ }\bibfield  {title}
  {\bibinfo {title} {Quantum antiferromagnet bluebellite comprising a
  maple-leaf lattice made of spin-1/2 {C}u$^{2+}$ ions},\ }\href
  {https://doi.org/10.1103/PhysRevB.104.174439} {\bibfield  {journal} {\bibinfo
   {journal} {Phys. Rev. B}\ }\textbf {\bibinfo {volume} {104}},\ \bibinfo
  {pages} {174439} (\bibinfo {year} {2021})}\BibitemShut {NoStop}%
\bibitem [{\citenamefont {Eisert}\ \emph {et~al.}(2010)\citenamefont {Eisert},
  \citenamefont {Cramer},\ and\ \citenamefont {Plenio}}]{Eisert2010}%
  \BibitemOpen
  \bibfield  {author} {\bibinfo {author} {\bibfnamefont {J.}~\bibnamefont
  {Eisert}}, \bibinfo {author} {\bibfnamefont {M.}~\bibnamefont {Cramer}},\
  and\ \bibinfo {author} {\bibfnamefont {M.~B.}\ \bibnamefont {Plenio}},\
  }\bibfield  {title} {\bibinfo {title} {Colloquium: Area laws for the
  entanglement entropy},\ }\href {https://doi.org/10.1103/RevModPhys.82.277}
  {\bibfield  {journal} {\bibinfo  {journal} {Rev. Mod. Phys.}\ }\textbf
  {\bibinfo {volume} {82}},\ \bibinfo {pages} {277} (\bibinfo {year}
  {2010})}\BibitemShut {NoStop}%
\bibitem [{\citenamefont {Orús}(2014)}]{Orus2014}%
  \BibitemOpen
  \bibfield  {author} {\bibinfo {author} {\bibfnamefont {R.}~\bibnamefont
  {Orús}},\ }\bibfield  {title} {\bibinfo {title} {{A practical introduction
  to tensor networks: Matrix product states and projected entangled pair
  states}},\ }\href {https://doi.org/https://doi.org/10.1016/j.aop.2014.06.013}
  {\bibfield  {journal} {\bibinfo  {journal} {Ann. Phys. (NY)}\ }\textbf
  {\bibinfo {volume} {349}},\ \bibinfo {pages} {117} (\bibinfo {year}
  {2014})}\BibitemShut {NoStop}%
\bibitem [{\citenamefont {Naumann}\ \emph {et~al.}(2024)\citenamefont
  {Naumann}, \citenamefont {Weerda}, \citenamefont {Rizzi}, \citenamefont
  {Eisert},\ and\ \citenamefont {Schmoll}}]{Naumann2023}%
  \BibitemOpen
  \bibfield  {author} {\bibinfo {author} {\bibfnamefont {J.}~\bibnamefont
  {Naumann}}, \bibinfo {author} {\bibfnamefont {E.~L.}\ \bibnamefont {Weerda}},
  \bibinfo {author} {\bibfnamefont {M.}~\bibnamefont {Rizzi}}, \bibinfo
  {author} {\bibfnamefont {J.}~\bibnamefont {Eisert}},\ and\ \bibinfo {author}
  {\bibfnamefont {P.}~\bibnamefont {Schmoll}},\ }\bibfield  {title} {\bibinfo
  {title} {{An introduction to infinite projected entangled-pair state methods
  for variational ground state simulations using automatic differentiation}},\
  }\href {https://doi.org/10.21468/SciPostPhysLectNotes.86} {\bibfield
  {journal} {\bibinfo  {journal} {SciPost Phys. Lect. Notes}\ ,\ \bibinfo
  {pages} {086}} (\bibinfo {year} {2024})}\BibitemShut {NoStop}%
\bibitem [{\citenamefont {Corboz}(2016)}]{Corboz2016}%
  \BibitemOpen
  \bibfield  {author} {\bibinfo {author} {\bibfnamefont {P.}~\bibnamefont
  {Corboz}},\ }\bibfield  {title} {\bibinfo {title} {Variational optimization
  with infinite projected entangled-pair states},\ }\href
  {https://doi.org/10.1103/PhysRevB.94.035133} {\bibfield  {journal} {\bibinfo
  {journal} {Phys. Rev. B}\ }\textbf {\bibinfo {volume} {94}},\ \bibinfo
  {pages} {035133} (\bibinfo {year} {2016})}\BibitemShut {NoStop}%
\bibitem [{\citenamefont {Ponsioen}\ and\ \citenamefont
  {Corboz}(2020)}]{Ponsioen2020}%
  \BibitemOpen
  \bibfield  {author} {\bibinfo {author} {\bibfnamefont {B.}~\bibnamefont
  {Ponsioen}}\ and\ \bibinfo {author} {\bibfnamefont {P.}~\bibnamefont
  {Corboz}},\ }\bibfield  {title} {\bibinfo {title} {Excitations with projected
  entangled pair states using the corner transfer matrix method},\ }\href
  {https://doi.org/10.1103/PhysRevB.101.195109} {\bibfield  {journal} {\bibinfo
   {journal} {Phys. Rev. B}\ }\textbf {\bibinfo {volume} {101}},\ \bibinfo
  {pages} {195109} (\bibinfo {year} {2020})}\BibitemShut {NoStop}%
\bibitem [{\citenamefont {Ponsioen}\ \emph {et~al.}(2023)\citenamefont
  {Ponsioen}, \citenamefont {Hasik},\ and\ \citenamefont
  {Corboz}}]{Ponsioen2023}%
  \BibitemOpen
  \bibfield  {author} {\bibinfo {author} {\bibfnamefont {B.}~\bibnamefont
  {Ponsioen}}, \bibinfo {author} {\bibfnamefont {J.}~\bibnamefont {Hasik}},\
  and\ \bibinfo {author} {\bibfnamefont {P.}~\bibnamefont {Corboz}},\
  }\bibfield  {title} {\bibinfo {title} {Improved summations of $n$-point
  correlation functions of projected entangled-pair states},\ }\href
  {https://doi.org/10.1103/PhysRevB.108.195111} {\bibfield  {journal} {\bibinfo
   {journal} {Phys. Rev. B}\ }\textbf {\bibinfo {volume} {108}},\ \bibinfo
  {pages} {195111} (\bibinfo {year} {2023})}\BibitemShut {NoStop}%
\bibitem [{\citenamefont {Haraldsen}\ \emph {et~al.}(2005)\citenamefont
  {Haraldsen}, \citenamefont {Barnes},\ and\ \citenamefont
  {Musfeldt}}]{Haraldsen-2005}%
  \BibitemOpen
  \bibfield  {author} {\bibinfo {author} {\bibfnamefont {J.~T.}\ \bibnamefont
  {Haraldsen}}, \bibinfo {author} {\bibfnamefont {T.}~\bibnamefont {Barnes}},\
  and\ \bibinfo {author} {\bibfnamefont {J.~L.}\ \bibnamefont {Musfeldt}},\
  }\bibfield  {title} {\bibinfo {title} {{Neutron scattering and magnetic
  observables for $S=1/2$ spin clusters and molecular magnets}},\ }\href
  {https://doi.org/10.1103/PhysRevB.71.064403} {\bibfield  {journal} {\bibinfo
  {journal} {Phys. Rev. B}\ }\textbf {\bibinfo {volume} {71}},\ \bibinfo
  {pages} {064403} (\bibinfo {year} {2005})}\BibitemShut {NoStop}%
\bibitem [{\citenamefont {Singh}\ and\ \citenamefont
  {Johnston}(2007)}]{Singh-2007}%
  \BibitemOpen
  \bibfield  {author} {\bibinfo {author} {\bibfnamefont {Y.}~\bibnamefont
  {Singh}}\ and\ \bibinfo {author} {\bibfnamefont {D.~C.}\ \bibnamefont
  {Johnston}},\ }\bibfield  {title} {\bibinfo {title} {{Singlet ground state in
  the spin-$\frac{1}{2}$ dimer compound
  ${\mathrm{Sr}}_{3}{\mathrm{Cr}}_{2}{\mathrm{O}}_{8}$}},\ }\href
  {https://doi.org/10.1103/PhysRevB.76.012407} {\bibfield  {journal} {\bibinfo
  {journal} {Phys. Rev. B}\ }\textbf {\bibinfo {volume} {76}},\ \bibinfo
  {pages} {012407} (\bibinfo {year} {2007})}\BibitemShut {NoStop}%
\bibitem [{\citenamefont {Quintero-Castro}\ \emph {et~al.}(2010)\citenamefont
  {Quintero-Castro}, \citenamefont {Lake}, \citenamefont {Wheeler},
  \citenamefont {Islam}, \citenamefont {Guidi}, \citenamefont {Rule},
  \citenamefont {Izaola}, \citenamefont {Russina}, \citenamefont {Kiefer},\
  and\ \citenamefont {Skourski}}]{Castro-2010}%
  \BibitemOpen
  \bibfield  {author} {\bibinfo {author} {\bibfnamefont {D.~L.}\ \bibnamefont
  {Quintero-Castro}}, \bibinfo {author} {\bibfnamefont {B.}~\bibnamefont
  {Lake}}, \bibinfo {author} {\bibfnamefont {E.~M.}\ \bibnamefont {Wheeler}},
  \bibinfo {author} {\bibfnamefont {A.~T. M.~N.}\ \bibnamefont {Islam}},
  \bibinfo {author} {\bibfnamefont {T.}~\bibnamefont {Guidi}}, \bibinfo
  {author} {\bibfnamefont {K.~C.}\ \bibnamefont {Rule}}, \bibinfo {author}
  {\bibfnamefont {Z.}~\bibnamefont {Izaola}}, \bibinfo {author} {\bibfnamefont
  {M.}~\bibnamefont {Russina}}, \bibinfo {author} {\bibfnamefont
  {K.}~\bibnamefont {Kiefer}},\ and\ \bibinfo {author} {\bibfnamefont
  {Y.}~\bibnamefont {Skourski}},\ }\bibfield  {title} {\bibinfo {title}
  {{Magnetic excitations of the gapped quantum spin dimer antiferromagnet
  ${\text{Sr}}_{3}{\text{Cr}}_{2}{\text{O}}_{8}$}},\ }\href
  {https://doi.org/10.1103/PhysRevB.81.014415} {\bibfield  {journal} {\bibinfo
  {journal} {Phys. Rev. B}\ }\textbf {\bibinfo {volume} {81}},\ \bibinfo
  {pages} {014415} (\bibinfo {year} {2010})}\BibitemShut {NoStop}%
\bibitem [{\citenamefont {Ghosh}\ \emph {et~al.}(2023)\citenamefont {Ghosh},
  \citenamefont {Seufert}, \citenamefont {M\"uller}, \citenamefont {Mila},\
  and\ \citenamefont {Thomale}}]{Ghosh-2023_mf}%
  \BibitemOpen
  \bibfield  {author} {\bibinfo {author} {\bibfnamefont {P.}~\bibnamefont
  {Ghosh}}, \bibinfo {author} {\bibfnamefont {J.}~\bibnamefont {Seufert}},
  \bibinfo {author} {\bibfnamefont {T.}~\bibnamefont {M\"uller}}, \bibinfo
  {author} {\bibfnamefont {F.}~\bibnamefont {Mila}},\ and\ \bibinfo {author}
  {\bibfnamefont {R.}~\bibnamefont {Thomale}},\ }\bibfield  {title} {\bibinfo
  {title} {{Maple leaf antiferromagnet in a magnetic field}},\ }\href
  {https://doi.org/10.1103/PhysRevB.108.L060406} {\bibfield  {journal}
  {\bibinfo  {journal} {Phys. Rev. B}\ }\textbf {\bibinfo {volume} {108}},\
  \bibinfo {pages} {L060406} (\bibinfo {year} {2023})}\BibitemShut {NoStop}%
\bibitem [{\citenamefont {Affleck}(1988)}]{Affleck-1988}%
  \BibitemOpen
  \bibfield  {author} {\bibinfo {author} {\bibfnamefont {I.}~\bibnamefont
  {Affleck}},\ }\bibfield  {title} {\bibinfo {title} {{Spin gap and symmetry
  breaking in ${\mathrm{CuO}}_{2}$ layers and other antiferromagnets}},\ }\href
  {https://doi.org/10.1103/PhysRevB.37.5186} {\bibfield  {journal} {\bibinfo
  {journal} {Phys. Rev. B}\ }\textbf {\bibinfo {volume} {37}},\ \bibinfo
  {pages} {5186} (\bibinfo {year} {1988})}\BibitemShut {NoStop}%
\bibitem [{\citenamefont {Oshikawa}\ \emph {et~al.}(1997)\citenamefont
  {Oshikawa}, \citenamefont {Yamanaka},\ and\ \citenamefont
  {Affleck}}]{Oshikawa-1997}%
  \BibitemOpen
  \bibfield  {author} {\bibinfo {author} {\bibfnamefont {M.}~\bibnamefont
  {Oshikawa}}, \bibinfo {author} {\bibfnamefont {M.}~\bibnamefont {Yamanaka}},\
  and\ \bibinfo {author} {\bibfnamefont {I.}~\bibnamefont {Affleck}},\
  }\bibfield  {title} {\bibinfo {title} {{Magnetization Plateaus in Spin
  Chains: ``Haldane Gap'' for Half-Integer Spins}},\ }\href
  {https://doi.org/10.1103/PhysRevLett.78.1984} {\bibfield  {journal} {\bibinfo
   {journal} {Phys. Rev. Lett.}\ }\textbf {\bibinfo {volume} {78}},\ \bibinfo
  {pages} {1984} (\bibinfo {year} {1997})}\BibitemShut {NoStop}%
\bibitem [{\citenamefont {Beck}\ \emph {et~al.}(2024)\citenamefont {Beck},
  \citenamefont {Bodky}, \citenamefont {Motruk}, \citenamefont {M\"uller},
  \citenamefont {Thomale},\ and\ \citenamefont {Ghosh}}]{Beck-2024}%
  \BibitemOpen
  \bibfield  {author} {\bibinfo {author} {\bibfnamefont {J.}~\bibnamefont
  {Beck}}, \bibinfo {author} {\bibfnamefont {J.}~\bibnamefont {Bodky}},
  \bibinfo {author} {\bibfnamefont {J.}~\bibnamefont {Motruk}}, \bibinfo
  {author} {\bibfnamefont {T.}~\bibnamefont {M\"uller}}, \bibinfo {author}
  {\bibfnamefont {R.}~\bibnamefont {Thomale}},\ and\ \bibinfo {author}
  {\bibfnamefont {P.}~\bibnamefont {Ghosh}},\ }\bibfield  {title} {\bibinfo
  {title} {{Phase diagram of the $J\text{\ensuremath{-}}{J}_{d}$ Heisenberg
  model on the maple leaf lattice: Neural networks and density matrix
  renormalization group}},\ }\href
  {https://doi.org/10.1103/PhysRevB.109.184422} {\bibfield  {journal} {\bibinfo
   {journal} {Phys. Rev. B}\ }\textbf {\bibinfo {volume} {109}},\ \bibinfo
  {pages} {184422} (\bibinfo {year} {2024})}\BibitemShut {NoStop}%
\bibitem [{\citenamefont {Gemb\'e}\ \emph {et~al.}(2024)\citenamefont
  {Gemb\'e}, \citenamefont {Gresista}, \citenamefont {Schmidt}, \citenamefont
  {Hickey}, \citenamefont {Iqbal},\ and\ \citenamefont {Trebst}}]{Gembe-2024}%
  \BibitemOpen
  \bibfield  {author} {\bibinfo {author} {\bibfnamefont {M.}~\bibnamefont
  {Gemb\'e}}, \bibinfo {author} {\bibfnamefont {L.}~\bibnamefont {Gresista}},
  \bibinfo {author} {\bibfnamefont {H.-J.}\ \bibnamefont {Schmidt}}, \bibinfo
  {author} {\bibfnamefont {C.}~\bibnamefont {Hickey}}, \bibinfo {author}
  {\bibfnamefont {Y.}~\bibnamefont {Iqbal}},\ and\ \bibinfo {author}
  {\bibfnamefont {S.}~\bibnamefont {Trebst}},\ }\bibfield  {title} {\bibinfo
  {title} {{Noncoplanar orders and quantum disordered states in maple-leaf
  antiferromagnets}},\ }\href {https://doi.org/10.1103/PhysRevB.110.085151}
  {\bibfield  {journal} {\bibinfo  {journal} {Phys. Rev. B}\ }\textbf {\bibinfo
  {volume} {110}},\ \bibinfo {pages} {085151} (\bibinfo {year}
  {2024})}\BibitemShut {NoStop}%
\bibitem [{\citenamefont {Ghosh}(2024)}]{ghosh2024spin}%
  \BibitemOpen
  \bibfield  {author} {\bibinfo {author} {\bibfnamefont {P.}~\bibnamefont
  {Ghosh}},\ }\bibfield  {title} {\bibinfo {title} {{Triplon analysis of
  magnetic disorder and order in maple-leaf Heisenberg magnet}},\ }\href
  {https://doi.org/10.1088/1361-648X/ad69f4} {\bibfield  {journal} {\bibinfo
  {journal} {J. Phys. Condens. Matter}\ }\textbf {\bibinfo {volume} {36}},\
  \bibinfo {pages} {455803} (\bibinfo {year} {2024})}\BibitemShut {NoStop}%
\bibitem [{\citenamefont {Sonnenschein}\ \emph {et~al.}(2024)\citenamefont
  {Sonnenschein}, \citenamefont {Maity}, \citenamefont {Liu}, \citenamefont
  {Thomale}, \citenamefont {Ferrari},\ and\ \citenamefont
  {Iqbal}}]{sonnenschein2024}%
  \BibitemOpen
  \bibfield  {author} {\bibinfo {author} {\bibfnamefont {J.}~\bibnamefont
  {Sonnenschein}}, \bibinfo {author} {\bibfnamefont {A.}~\bibnamefont {Maity}},
  \bibinfo {author} {\bibfnamefont {C.}~\bibnamefont {Liu}}, \bibinfo {author}
  {\bibfnamefont {R.}~\bibnamefont {Thomale}}, \bibinfo {author} {\bibfnamefont
  {F.}~\bibnamefont {Ferrari}},\ and\ \bibinfo {author} {\bibfnamefont
  {Y.}~\bibnamefont {Iqbal}},\ }\bibfield  {title} {\bibinfo {title}
  {{Candidate quantum spin liquids on the maple-leaf lattice}},\ }\href
  {https://doi.org/10.1103/PhysRevB.110.014414} {\bibfield  {journal} {\bibinfo
   {journal} {Phys. Rev. B}\ }\textbf {\bibinfo {volume} {110}},\ \bibinfo
  {pages} {014414} (\bibinfo {year} {2024})}\BibitemShut {NoStop}%
\bibitem [{\citenamefont {Liechtenstein}\ \emph {et~al.}(1995)\citenamefont
  {Liechtenstein}, \citenamefont {Anisimov},\ and\ \citenamefont
  {Zaanen}}]{Liechtenstein1995}%
  \BibitemOpen
  \bibfield  {author} {\bibinfo {author} {\bibfnamefont {A.~I.}\ \bibnamefont
  {Liechtenstein}}, \bibinfo {author} {\bibfnamefont {V.~I.}\ \bibnamefont
  {Anisimov}},\ and\ \bibinfo {author} {\bibfnamefont {J.}~\bibnamefont
  {Zaanen}},\ }\bibfield  {title} {\bibinfo {title} {{Density-functional theory
  and strong interactions: Orbital ordering in Mott-Hubbard insulators}},\
  }\href {https://doi.org/10.1103/PhysRevB.52.R5467} {\bibfield  {journal}
  {\bibinfo  {journal} {Phys. Rev. B}\ }\textbf {\bibinfo {volume} {52}},\
  \bibinfo {pages} {R5467} (\bibinfo {year} {1995})}\BibitemShut {NoStop}%
\bibitem [{\citenamefont {Pohle}\ and\ \citenamefont
  {Jaubert}(2023)}]{Pohle2023}%
  \BibitemOpen
  \bibfield  {author} {\bibinfo {author} {\bibfnamefont {R.}~\bibnamefont
  {Pohle}}\ and\ \bibinfo {author} {\bibfnamefont {L.~D.~C.}\ \bibnamefont
  {Jaubert}},\ }\bibfield  {title} {\bibinfo {title} {Curie-law crossover in
  spin liquids},\ }\href {https://doi.org/10.1103/PhysRevB.108.024411}
  {\bibfield  {journal} {\bibinfo  {journal} {Phys. Rev. B}\ }\textbf {\bibinfo
  {volume} {108}},\ \bibinfo {pages} {024411} (\bibinfo {year}
  {2023})}\BibitemShut {NoStop}%
\bibitem [{\citenamefont {Gonzalez}\ \emph {et~al.}(2025)\citenamefont
  {Gonzalez}, \citenamefont {Iqbal}, \citenamefont {Reuther},\ and\
  \citenamefont {Jeschke}}]{Gonzalez2025}%
  \BibitemOpen
  \bibfield  {author} {\bibinfo {author} {\bibfnamefont {M.~G.}\ \bibnamefont
  {Gonzalez}}, \bibinfo {author} {\bibfnamefont {Y.}~\bibnamefont {Iqbal}},
  \bibinfo {author} {\bibfnamefont {J.}~\bibnamefont {Reuther}},\ and\ \bibinfo
  {author} {\bibfnamefont {H.~O.}\ \bibnamefont {Jeschke}},\ }\bibfield
  {title} {\bibinfo {title} {Field-induced spin liquid in the decorated
  square-kagome antiferromagnet nabokoite \ce{KCu7TeO4(SO4)5Cl}},\ }\href
  {https://doi.org/10.1038/s43246-025-00806-2} {\bibfield  {journal} {\bibinfo
  {journal} {Comm. Mater.}\ }\textbf {\bibinfo {volume} {6}},\ \bibinfo {pages}
  {96} (\bibinfo {year} {2025})}\BibitemShut {NoStop}%
\bibitem [{\citenamefont {Schmoll}\ \emph {et~al.}(2023)\citenamefont
  {Schmoll}, \citenamefont {Kshetrimayum}, \citenamefont {Naumann},
  \citenamefont {Eisert},\ and\ \citenamefont {Iqbal}}]{Schmoll-2023}%
  \BibitemOpen
  \bibfield  {author} {\bibinfo {author} {\bibfnamefont {P.}~\bibnamefont
  {Schmoll}}, \bibinfo {author} {\bibfnamefont {A.}~\bibnamefont
  {Kshetrimayum}}, \bibinfo {author} {\bibfnamefont {J.}~\bibnamefont
  {Naumann}}, \bibinfo {author} {\bibfnamefont {J.}~\bibnamefont {Eisert}},\
  and\ \bibinfo {author} {\bibfnamefont {Y.}~\bibnamefont {Iqbal}},\ }\bibfield
   {title} {\bibinfo {title} {{Tensor network study of the spin-$\frac{1}{2}$
  Heisenberg antiferromagnet on the shuriken lattice}},\ }\href
  {https://doi.org/10.1103/PhysRevB.107.064406} {\bibfield  {journal} {\bibinfo
   {journal} {Phys. Rev. B}\ }\textbf {\bibinfo {volume} {107}},\ \bibinfo
  {pages} {064406} (\bibinfo {year} {2023})}\BibitemShut {NoStop}%
\bibitem [{\citenamefont {Nishino}\ and\ \citenamefont
  {Okunishi}(1996)}]{Nishino1996}%
  \BibitemOpen
  \bibfield  {author} {\bibinfo {author} {\bibfnamefont {T.}~\bibnamefont
  {Nishino}}\ and\ \bibinfo {author} {\bibfnamefont {K.}~\bibnamefont
  {Okunishi}},\ }\bibfield  {title} {\bibinfo {title} {{Corner Transfer Matrix
  Renormalization Group Method}},\ }\href {https://doi.org/10.1143/JPSJ.65.891}
  {\bibfield  {journal} {\bibinfo  {journal} {J. Phys. Soc. Jpn.}\ }\textbf
  {\bibinfo {volume} {65}},\ \bibinfo {pages} {891} (\bibinfo {year}
  {1996})}\BibitemShut {NoStop}%
\bibitem [{\citenamefont {Nishino}\ and\ \citenamefont
  {Okunishi}(1997)}]{Nishino1997}%
  \BibitemOpen
  \bibfield  {author} {\bibinfo {author} {\bibfnamefont {T.}~\bibnamefont
  {Nishino}}\ and\ \bibinfo {author} {\bibfnamefont {K.}~\bibnamefont
  {Okunishi}},\ }\bibfield  {title} {\bibinfo {title} {{Corner Transfer Matrix
  Algorithm for Classical Renormalization Group}},\ }\href
  {https://doi.org/10.1143/JPSJ.66.3040} {\bibfield  {journal} {\bibinfo
  {journal} {J. Phys. Soc. Jpn.}\ }\textbf {\bibinfo {volume} {66}},\ \bibinfo
  {pages} {3040} (\bibinfo {year} {1997})}\BibitemShut {NoStop}%
\bibitem [{\citenamefont {Or\'us}\ and\ \citenamefont
  {Vidal}(2009)}]{Orus2009}%
  \BibitemOpen
  \bibfield  {author} {\bibinfo {author} {\bibfnamefont {R.}~\bibnamefont
  {Or\'us}}\ and\ \bibinfo {author} {\bibfnamefont {G.}~\bibnamefont {Vidal}},\
  }\bibfield  {title} {\bibinfo {title} {Simulation of two-dimensional quantum
  systems on an infinite lattice revisited: Corner transfer matrix for tensor
  contraction},\ }\href {https://doi.org/10.1103/PhysRevB.80.094403} {\bibfield
   {journal} {\bibinfo  {journal} {Phys. Rev. B}\ }\textbf {\bibinfo {volume}
  {80}},\ \bibinfo {pages} {094403} (\bibinfo {year} {2009})}\BibitemShut
  {NoStop}%
\bibitem [{\citenamefont {Jordan}\ \emph {et~al.}(2008)\citenamefont {Jordan},
  \citenamefont {Or\'us}, \citenamefont {Vidal}, \citenamefont {Verstraete},\
  and\ \citenamefont {Cirac}}]{Jordan2008}%
  \BibitemOpen
  \bibfield  {author} {\bibinfo {author} {\bibfnamefont {J.}~\bibnamefont
  {Jordan}}, \bibinfo {author} {\bibfnamefont {R.}~\bibnamefont {Or\'us}},
  \bibinfo {author} {\bibfnamefont {G.}~\bibnamefont {Vidal}}, \bibinfo
  {author} {\bibfnamefont {F.}~\bibnamefont {Verstraete}},\ and\ \bibinfo
  {author} {\bibfnamefont {J.~I.}\ \bibnamefont {Cirac}},\ }\bibfield  {title}
  {\bibinfo {title} {Classical simulation of infinite-size quantum lattice
  systems in two spatial dimensions},\ }\href
  {https://doi.org/10.1103/PhysRevLett.101.250602} {\bibfield  {journal}
  {\bibinfo  {journal} {Phys. Rev. Lett.}\ }\textbf {\bibinfo {volume} {101}},\
  \bibinfo {pages} {250602} (\bibinfo {year} {2008})}\BibitemShut {NoStop}%
\bibitem [{\citenamefont {Kshetrimayum}\ \emph {et~al.}(2019)\citenamefont
  {Kshetrimayum}, \citenamefont {Rizzi}, \citenamefont {Eisert},\ and\
  \citenamefont {Or\'us}}]{Kshetrimayum2019}%
  \BibitemOpen
  \bibfield  {author} {\bibinfo {author} {\bibfnamefont {A.}~\bibnamefont
  {Kshetrimayum}}, \bibinfo {author} {\bibfnamefont {M.}~\bibnamefont {Rizzi}},
  \bibinfo {author} {\bibfnamefont {J.}~\bibnamefont {Eisert}},\ and\ \bibinfo
  {author} {\bibfnamefont {R.}~\bibnamefont {Or\'us}},\ }\bibfield  {title}
  {\bibinfo {title} {Tensor network annealing algorithm for two-dimensional
  thermal states},\ }\href {https://doi.org/10.1103/PhysRevLett.122.070502}
  {\bibfield  {journal} {\bibinfo  {journal} {Phys. Rev. Lett.}\ }\textbf
  {\bibinfo {volume} {122}},\ \bibinfo {pages} {070502} (\bibinfo {year}
  {2019})}\BibitemShut {NoStop}%
\bibitem [{\citenamefont {Schmoll}\ \emph {et~al.}(2025)\citenamefont
  {Schmoll}, \citenamefont {Jeschke},\ and\ \citenamefont
  {Iqbal}}]{SpangoliteData}%
  \BibitemOpen
  \bibfield  {author} {\bibinfo {author} {\bibfnamefont {P.}~\bibnamefont
  {Schmoll}}, \bibinfo {author} {\bibfnamefont {H.}~\bibnamefont {Jeschke}},\
  and\ \bibinfo {author} {\bibfnamefont {Y.}~\bibnamefont {Iqbal}},\ }\href
  {https://doi.org/10.5281/zenodo.16085808} {\bibinfo {title} {{Data for
  ``Tensor network analysis of the maple-leaf antiferromagnet spangolite''}}},\
  \bibinfo {howpublished} {Zenodo} (\bibinfo {year} {2025})\BibitemShut
  {NoStop}%
\bibitem [{\citenamefont {Haegeman}\ \emph {et~al.}(2025)\citenamefont
  {Haegeman}, \citenamefont {Devos},\ and\ \citenamefont
  {contributors}}]{TensorKit.jl}%
  \BibitemOpen
  \bibfield  {author} {\bibinfo {author} {\bibfnamefont {J.}~\bibnamefont
  {Haegeman}}, \bibinfo {author} {\bibfnamefont {L.}~\bibnamefont {Devos}},\
  and\ \bibinfo {author} {\bibnamefont {contributors}},\ }\href
  {https://doi.org/10.5281/zenodo.15880132} {\bibinfo {title}
  {{Jutho/TensorKit.jl: v0.14.8}}},\ \bibinfo {howpublished} {Zenodo} (\bibinfo
  {year} {2025})\BibitemShut {NoStop}%
\bibitem [{\citenamefont {Schmoll}(2025)}]{SpangoliteCode}%
  \BibitemOpen
  \bibfield  {author} {\bibinfo {author} {\bibfnamefont {P.}~\bibnamefont
  {Schmoll}},\ }\href {https://github.com/philihps/Spangolite} {\bibinfo
  {title} {{Spangolite simulation code}}},\ \bibinfo {howpublished} {GitHub}
  (\bibinfo {year} {2025})\BibitemShut {NoStop}%
\bibitem [{\citenamefont {Bennett}\ \emph {et~al.}(2020)\citenamefont
  {Bennett}, \citenamefont {Melchers},\ and\ \citenamefont
  {Proppe}}]{Bennett2020}%
  \BibitemOpen
  \bibfield  {author} {\bibinfo {author} {\bibfnamefont {L.}~\bibnamefont
  {Bennett}}, \bibinfo {author} {\bibfnamefont {B.}~\bibnamefont {Melchers}},\
  and\ \bibinfo {author} {\bibfnamefont {B.}~\bibnamefont {Proppe}},\ }\href
  {http://dx.doi.org/10.17169/refubium-26754} {\bibinfo {title} {{Curta: A
  general-purpose high-performance computer at ZEDAT, Freie Universit{\"a}t
  Berlin}}} (\bibinfo {year} {2020})\BibitemShut {NoStop}%
\end{thebibliography}

%

\clearpage


\section*{Supplementary material}

the simple update and Trotterization of the Hamiltonian. Starting from an infinite-temperature state $\rho(\beta = 0)$, where the thermal density matrix is simply a tensor product of local identity matrices, the system is cooled down to the desired final temperature $\beta$. This procedure is implemented by successive cooling steps with infinitesimal temperature $\delta\beta$ to keep the overall errors controllable. This can be done in a similar fashion as in the simple update ground state optimization and it is visualized in Fig. 2 c. Again, after the evolution with the gate $e^{-\delta\beta H}$, the resulting iPESO network is truncated to the bulk bond dimension $\chi_B$. The accumulated truncation error $\varepsilon$ of the full procedure can be used to probe the accuracy of the simulations [22].

### 3. Calculation of expectation values

In order to compute expectation values, the honeycomb structure of the iPESS/iPESO ansatz can be mapped to a regular square lattice with lattice vectors as shown in Fig. 1 a. The resulting tensor network on the square lattice is known as an *infinite projected entangled pair state* (iPEPS) or *infinite projected entangled pair operator* (iPEPO) respectively, which includes six spins of the original MLL on every lattice site. This additional step is required to compute effective fixed-point environments of the infinite MLL, e.g. by a regular *corner transfer matrix renormalization group* (CTMRG) procedure, as shown in Fig. 2 d. Expectation values on the MLL can then be accurately computed by the evaluation of single-site and three-site expectation values on the coarse-grained square lattice, corresponding to the two different types of simplex configurations in the iPESS/iPESO ansatz. The approximations in the contraction of the infinite 2D lattice are controlled by a bond dimension for the effective environment tensors, denoted by $\chi_E$. In practise, this parameter has to be large enough so that expectation values are well converged. Besides taking the full CTMRG environment into account for the calculation of expectation values, it is also possible to only use a mean-field (MF) environment. This environment is readily available from the simple update for both ground and thermal states. MF expectation values are computationally much cheaper and have to be used for large bond dimensions due to the high computational cost of the CTMRG routine.



\end{document}